\documentclass[5p,fleqn]{elsarticle}
\usepackage[latin1]{inputenc}
\usepackage{graphicx}
\usepackage{hyperref}
\usepackage{upgreek}
\usepackage{lineno}

 \usepackage{epsfig} 
 \usepackage{float} 
 \usepackage[pdftex]{color} 
 \usepackage{amsmath} 
 \usepackage{amsfonts}

\renewcommand{\[}{\begin{equation}}
\renewcommand{\]}{\end{equation}}

\journal{Nuclear Instruments and Methods in Physics Research A}

\begin{document}

\title{Simulation and Experimental Study of Plasma Effects in Planar Silicon Sensors}

\author[hh]{Julian Becker}
\ead{Julian.Becker@desy.de}
\author[wias]{Klaus G\"{a}rtner}
\author[hh]{Robert Klanner}
\author[hll]{Rainer Richter}
\address[hh]{Institut f\"{u}r Experimentalphysik, Universit\"{a}t Hamburg,\\ Luruper Chaussee 149, 22761 Hamburg, Germany}
\address[wias]{Weierstra\ss-Institut f\"{u}r Angewandte Analysis und Stochastik, \\ Mohrenstr. 39, 10117 Berlin, Germany}
\address[hll]{Max-Planck-Institut Halbleiterlabor, \\ Otto-Hahn-Ring 6, 81739 M\"{u}nchen, Germany}

\begin{abstract}
In silicon sensors high densities of electron-hole pairs result in a change of the current pulse shape and spatial distribution of the collected charge compared to the situation in presence of low charge carrier densities.
This paper presents a detailed comparison of numerical simulations with time resolved current measurements on planar silicon sensors using 660~nm laser light to create different densities of electron hole pairs. 

\end{abstract}


\maketitle

\linenumbers
\setlength{\mathindent}{0mm}

\section{Introduction}
Silicon sensors are frequently used for the detection of radiation. These sensors are built as p-n junctions operated under reverse bias. Electron hole pairs are created by ionization or direct excitation and induce a current in the electrodes when drifting in the electric field until they reach a contact. 
When the created charge carrier densities are sufficiently high to modify the electric field in the sensor, significant changes, compared to the situation with low charge carrier densities, are observed (so called plasma effects). 
These effects have been observed in the detection of heavily ionizing particles and with high intensity laser light. 

Plasma effects are also expected for experiments at x-ray free electron lasers. The studies presented here aim at a quantitative understanding of the plasma effects for experiments at the European XFEL \cite{xfel}.

For high charge carrier densities the electrons and holes form a so called plasma, which dissolves slowly. The plasma boundaries effectively shield its inner region from the external electric field created by the external bias, thus altering the induced current pulse and increasing the charge collection time \cite{seibt}. Plasma effects decrease as the electric field increases \cite{williams}. Using incident ions of different masses and energies, the influence of material properties on plasma effects has been studied in detail in \cite{bohne}.

Electrostatic repulsion effects result in an increased lateral spread of the collected charge and thus in increased charge sharing between pixels, as shown for $\upalpha$-particles in \cite{campbell}. The effects on silicon sensors for the European XFEL were investigated using a focused high intensity laser to simulate x-rays in \cite{becker, phd}.

In this work, simulation results are compared to measured current pulses showing plasma distortions after illumination with focused laser light of high intensity.

\section{Experimental setup}

Charge carriers were created with a laser of 660~nm wavelength and the time resolved current pulses of the investigated diode were read out by a Miteq AM-1309 wideband amplifier and a Tektronix DPO 7254 2.5~GHz oscilloscope. 

The sample was mounted on a substrate that allowed light injection from both sides while providing a stable ($\pm$~0.1~K, rms) temperature in the range of 240~K to 340~K and applying the high voltage to the rear side of the diode.

The systematic error of the determination of the number of generated electron hole pairs was estimated to be below 2\% by injecting a defined charge into the readout system utilizing a defined test capacitance and a voltage step function.

\subsection{Equivalent circuit for SPICE simulations}

\begin{figure}[t!]
  \includegraphics[width=0.5\textwidth]{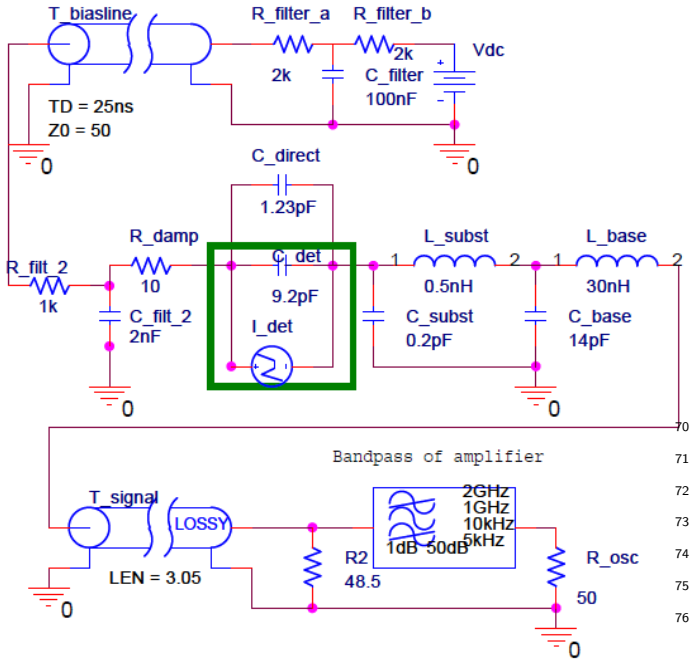}
  \centering
  \caption{Equivalent circuit for SPICE simulations. Dominant elements are the diode capacitance in parallel to the current source (marked by the rectangle), the system inductance (30~nH), the lossy transmission line (3.05~m RG-174 cable) and the band pass filtering properties of the amplifier.\label{circuit}}
\end{figure}

\begin{figure}[tb!]
  \includegraphics[width=0.5\textwidth]{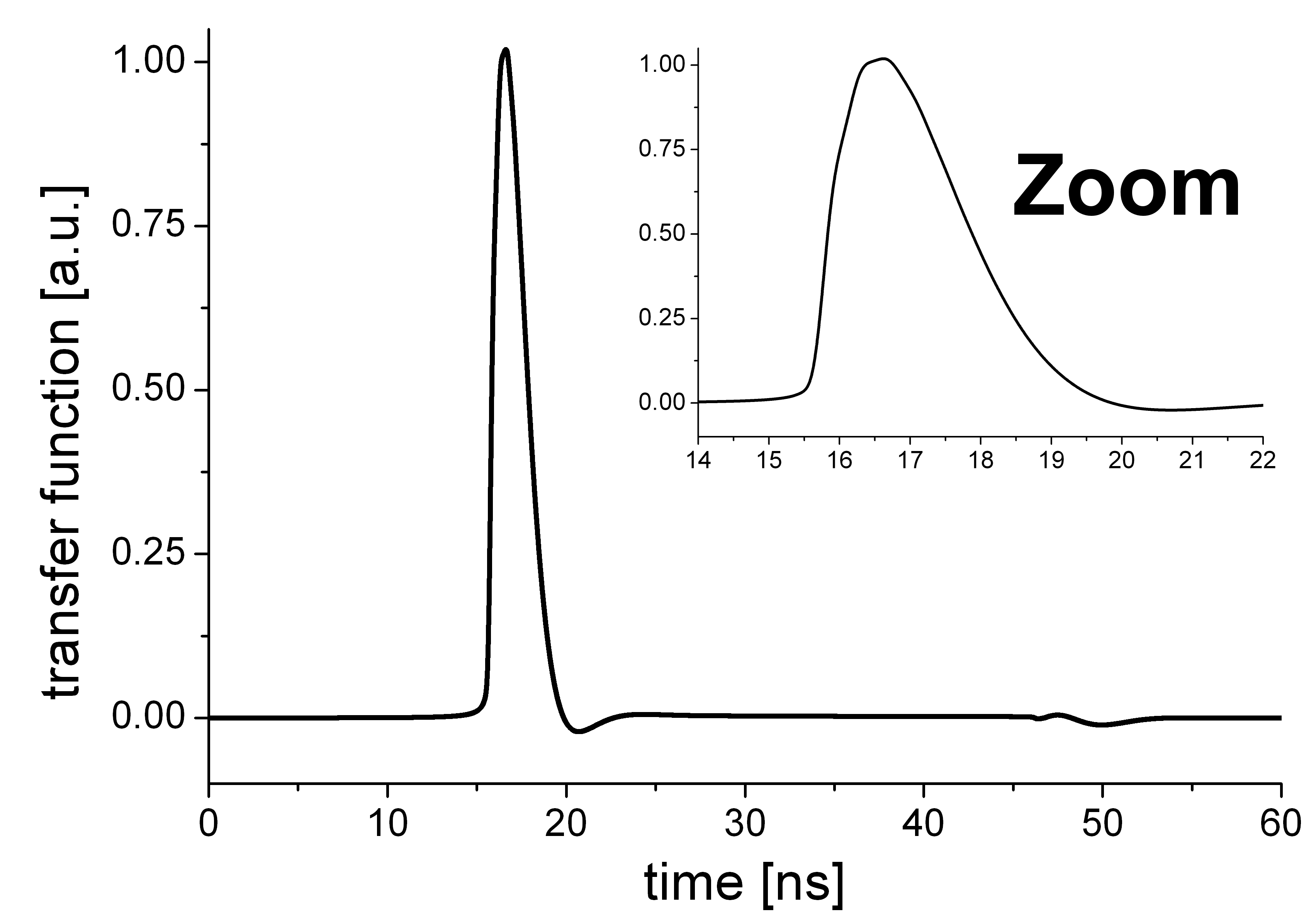}
  \centering
  \caption{Transfer function derived from the SPICE simulations for 9.2~pF capacitance. The inset shows a zoom of the peak structure of the transfer function using the same units on the x- and y-axis.\label{transfer}}
\end{figure}

An equivalent circuit of the setup was used for SPICE simulations (see Figure~\ref{circuit}). The equivalent circuit takes detector capacitance (9.2~pF), signal cable length (3.05~m), bandwidth of the amplifier (1~GHz) and imperfections of the setup (inductances and capacitances) into account. The transfer function of the whole setup has been checked and was found to introduce distortions. Thus all simulations have been convoluted with the transfer function (shown in Figure~\ref{transfer}).

\subsection{Laser properties}

The laser system \cite{pq} emits short and intense light pulses with a FWHM $<$~100~ps. The time resolved pulse structure was specified by the manufacturer. For this study laser light with a wavelength of 660~nm ($\pm$~2~nm) was used. As the pulse structure depends on the pulse energy, a constant energy of 140~pJ was chosen and attenuated with optical attenuators, which have no effect on the time structure of the pulses. 660~nm light has an attenuation length of roughly 3~$\upmu$m in silicon at 20$^\circ$C.

The laser beam was focused to a spot with a Gaussian profile with $\sigma$~$=$~10~$\upmu$m. In air the Rayleigh length (distance from focal point to the point where the beam radius increases by $\sqrt{2}$) is approximately 90~$\upmu$m. The high index of refraction of silicon ($\approx$~3.6) allows to assume a constant lateral beam profile for the entire absorption path.

\subsection{Investigated diode}

\begin{figure}[tb!]
  \includegraphics[width=0.5\textwidth]{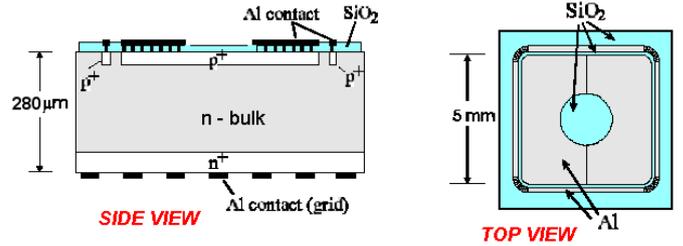}
  \centering
  \caption{Geometry of the investigated sample. The diode has a thickness of 280~$\upmu$m and allows injection of laser light from the junction side and opposite to the junction.\label{diode}}
\end{figure}

The investigated diode was a planar p$^+$nn$^+$ diode fabricated by CiS \cite{cis}. The silicon used is high resistivity n-type diffusion oxygenated float zone silicon with $<$100$>$ orientation manufactured by Siltronic \cite{siltronic}.
The effective doping of the sample was calculated from a capacitance measurement as function of voltage and is 8.2$\times$10$^{11}$~cm$^{-3}$. The resulting depletion voltage is 49~V with a dark current at the depletion voltage of 0.72~nA.
The sample has a very low concentration of lattice defects, leading to life times in the order of milliseconds, which is many orders of magnitude larger than the longest pulse duration recorded in this work. Any trapping effects have been considered negligible.

From the measured pad capacitance of 9.23~pF and the pad area of (4.95~mm)$^2$ we obtain a thickness of 280~$\upmu$m using the standard formula for a parallel plate capacitor without edge effects. For the measurement of the pad capacitance the capacitance of the guard ring to the backplane and the sensor edge is subtracted by the zero adjustment of the capacitance bridge. The remaining effect of the guard ring is estimated to be well below 1~\%. The estimated uncertainty of the diode thickness is $\pm$ 2~$\upmu$m.

Mechanical measurements of the thickness yield the somewhat higher value of 287~$\upmu$m but include 'dead' layers like implantations, passivations and aluminizations. 

The diode has an opening on the p$^+$ side and an aluminum grid on the n$^+$ side to allow light injection. The gap between the metalization of diode guard ring is 20~$\upmu$m wide, the distance between the corresponding implantations is 10~$\upmu$m. A sketch of the diode is shown in Figure~\ref{diode}.

\section{Simulations}

For the transport simulations the classical van Roosbroeck equations are used \cite{roosbroeck}. 

The lifetime of the plasma cloud is determined by the emission of carriers into the surrounding volume with low charge carrier density. A slow movement of the center of the plasma cloud results from the different probability to emit an electron or a hole from different parts of the plasma, depending on the distance to the closest electrode. 
Most of the charge carriers outside of the plasma are transported in the depleted volume. The electric field in this volume (outside of the plasma) is defined by the bulk doping and the external voltage.
Dominating influences in the simulation are the field induced mobility reduction, the initial cloud size and density.

\subsection{Simulated physics}

The fundamental equations describing the problem are the Poisson equation and the continuity equations for electrons and holes. 

\begin{eqnarray}
-\nabla \cdot \epsilon \nabla w + n - p &=& C\\
\frac{\partial n}{\partial t} + \nabla \cdot \mu_n n \nabla \phi_n &=& R \\ 
\frac{\partial p}{\partial t} - \nabla \cdot \mu_p p \nabla \phi_p &=& R 
\end{eqnarray} 

Where $\epsilon=\epsilon_0 \epsilon_r$ is the dielectric permittivity, $w$ the electrostatic potential, $n$ and $p$ the electron and hole density, $C$ the density of impurities, $\phi_{n,p}$ the quasi-Fermi potential for electrons or holes,  
$\mu_{n,p}$ the charge carrier mobilities and $R$ the recombination or generation rate.

Potentials are normalized to a constant reference potential and densities to a reference density. The meaning of the variables and parameters used are listed in the appendix.

The mobility models used are summarized and discussed in \cite{selberherr84}.
The mobility $\mu$ depends on lattice ($L$), ionized ($I$) and unionized ($N$) impurity scattering, carrier-carrier scattering ($np$) and the electric field, where ($I$), ($N$) and ($np$) have the meaning of densities.

Lattice and ionized impurity scattering is modeled using:
\begin{eqnarray}
	\mu_{n,p}^L &=& \mu_{n,p}^0 (\frac{T}{300K})^{-\alpha_{n,p}} \\
	\mu^{LI}_{n,p} &=& \frac{\mu_{n,p}^{L}}{\sqrt{1 + \frac{I}{C_{n,p}^{ref}+\frac{I}{S_{n,p}}}}}
\end{eqnarray}

Unionized impurity scattering is modeled using:
\begin{eqnarray}
  \mu^{N}(T) &=& \frac{0.041 q m^*_{n,p}}{N a_{Bohr} \hbar m_0 \epsilon_r } 
\left (\frac{2}{3}\sqrt{\frac{k_B T}{EN_{n,p}}}+\frac{1}{3}\sqrt{\frac{EN_{n,p}}{k_B T}} \right ) \\ 
  EN_{n,p} &=& 0.71 eV \frac{m^*_{n,p}}{m_0} \left ( \frac{\epsilon_0}{\epsilon}\right )^2 
\end{eqnarray}

Carrier-carrier scattering is implemented using the structure of Adler's model \cite{adler}:
\[
\mu^{np}=\frac{\mu_{Adl1}}{\sqrt{np} \ln(1+\mu_{Adl2}(np)^{-1/3})}
\]

The total mobility is approximated by using Mat\-thies\-sen's rule \cite{matt} to combine the contributions (they have the meaning of scaled inverse macroscopic cross 
sections $\Sigma$, the total cross-section $\Sigma_{tot}$ is the sum over all 
$\Sigma_i$ of uncorrelated scattering events $i$).

\[
1/\mu^{LINnp}=1/\mu^{LI} + 1/\mu^{N} + 1/\mu^{np}
\]
 
The total mobility is reduced by velocity saturation for high values of $|\nabla \phi_{n,p}|$. 
\begin{equation}\label{equ-velocity-sat}
\mu^{LINnpE}_{n,p}= \frac{\mu^{LINnp}_{n,p}}{(1+(\frac{\mu^{LINnp}_{n,p} |\nabla \phi_{n,p}|}{v_{sat}})^\beta)^{\frac{1}{\beta}}}
\end{equation}

In the simulations the driving force ($\nabla \phi_{n,p}$) is used to derive the mobility reduction\footnote{In the depleted bulk silicon this is identical to the derivation using the electric field ($\nabla w$).}. For the sake of simplicity we call this reduction 'field induced'.

\subsection{Numerical methods}

\begin{figure*}[t!]
  $\begin{array}{cc}
  \includegraphics[width=0.25\textwidth]{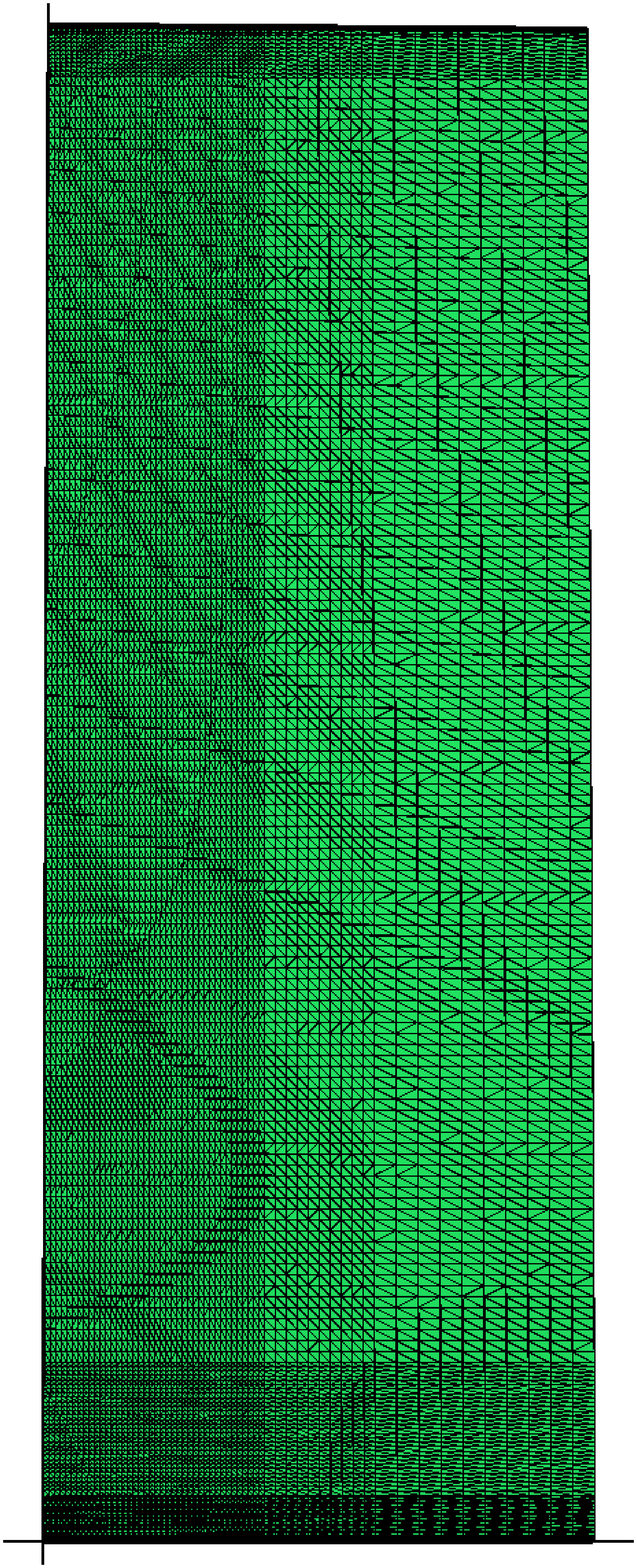} &
 	\includegraphics[width=0.75\textwidth]{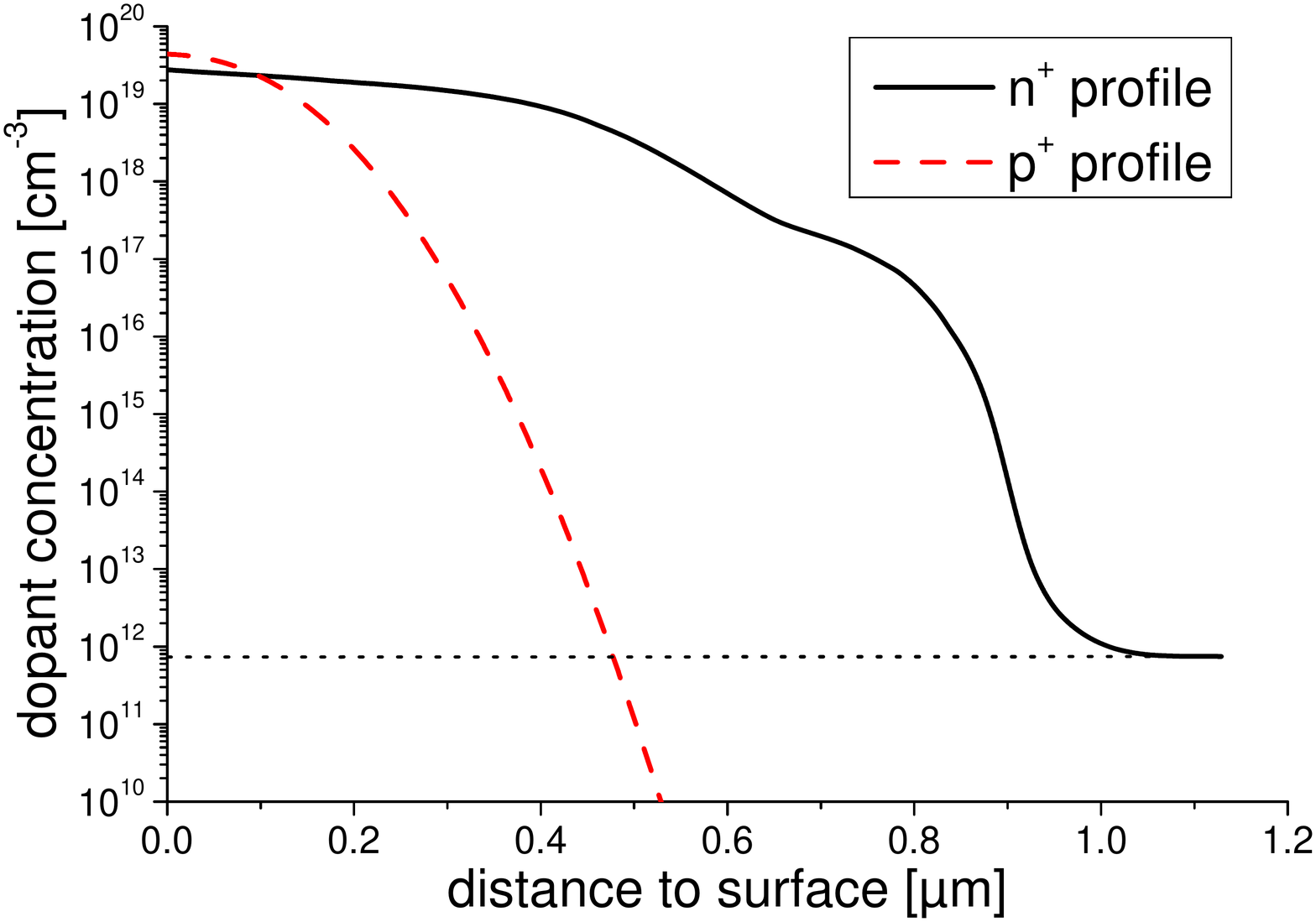} 
\end{array}$
  \centering
  \caption{The left graph displays a grid of mesh size $h$ defining the slice of a cylinder. Step sizes are $h_r=1,...,4\upmu m$ and $h_z=0.2,...,2 \upmu m$. Top and bottom of the grid are defined as ohmic contacts, the boundary conditions create a cylindrical symmetry. The right graph displays doping profiles as function of depth. The tail of the n$^+$ profile reflects the bulk doping concentration which is marked by the dotted horizontal line.\label{fig-diode-gridA1}}
\end{figure*}

The code follows the theory described in \cite{kg09}. 
The time integration is based on variable order implicit backward differentiation formulas (BDF)
with time step size and order control (see appendix).

For certain detector applications a very precise number of electron hole pairs (relative error $<<$ 10$^{-5}$) has to be generated in order to estimate possible charge losses in the device. In order to achieve this high precision an automated rescaling procedure is used.

The rescaling procedure uses a comparison of two integrations. For this, all external generation processes (sources, e.g. laser excitation) are parameterized in space and time. The first integration is done independent from the simulation process and results in a high precision source integral and the number of electron hole pairs that should be generated.

The second integration is executed during the simulation and uses the discrete time and space steps of the simulation for integration. The integration result is the number of generated electron hole pairs in the simulation. There is a small difference ($<$1\%) between generated and intended number of charge carriers due to the finite step sizes and time discretization errors.

Charge carrier numbers are rescaled to the intended numbers, derived values like current are rescaled as well. In this way the required high precision generation is assured.

The contact currents are evaluated by using test functions, which 
approximate the solutions of the related adjoint problem \cite{kg07}. 
These techniques are necessary to fulfill charge conservation requirements 
expected in detector design applications \cite{rg06}.

\section{Simulation parameters} 
A rather simple spatial domain is used to test the numerical methods. A 10$^\circ$ slice of a cylinder of 280~${\rm \upmu m}$ height and 100~${\rm \upmu m}$ radius is discretized by a set of tensor product grids that are rotated and split in tetrahedra to obtain a slightly anisotropic Delaunay grid. This allows
to align the edges in the main field direction and results in less
points compared to isotropic grids.
Thus nested grids, labeled $h$ (22990 nodes), $h/2$ (91339 nodes)and $h/4$ (364117 nodes) are possible (see Figure~\ref{fig-diode-gridA1}).

\begin{figure*}[t!]
  $\begin{array}{cc}
  \includegraphics[width=0.5\textwidth]{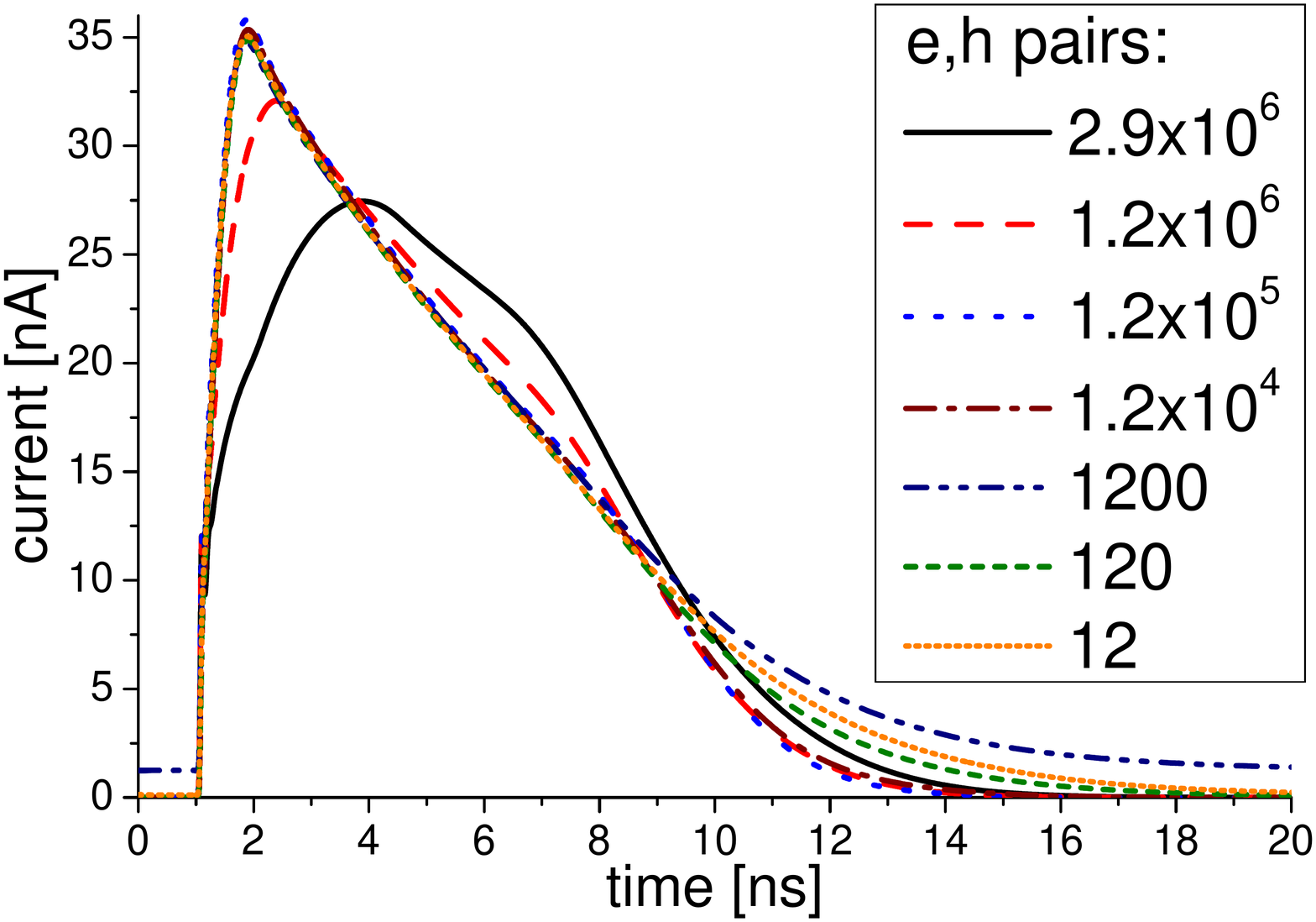} &
 	\includegraphics[width=0.5\textwidth]{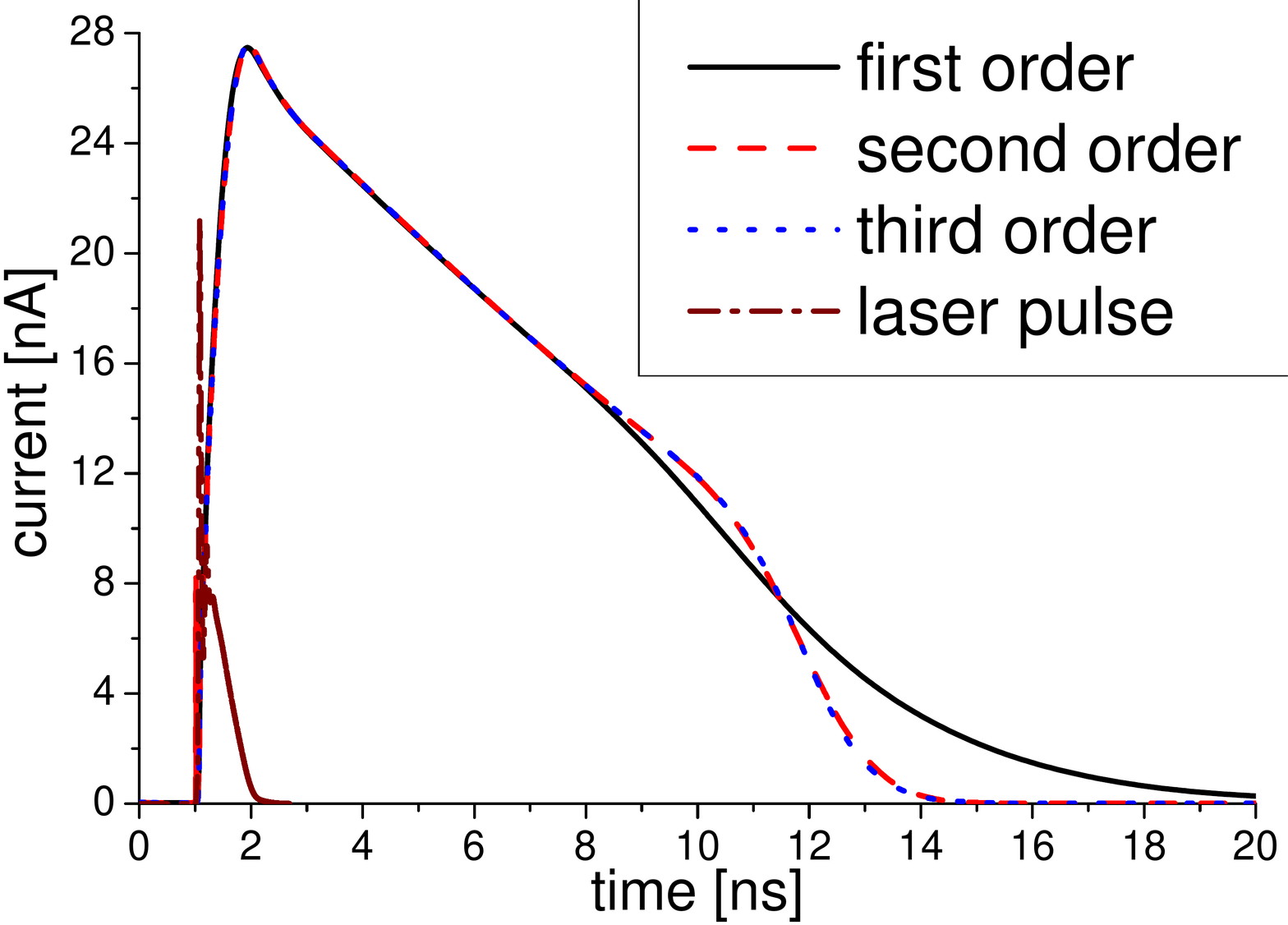} 
\end{array}$
  \centering
  \caption{The left graph shows the current pulse for different numbers of charge carriers created at the junction side, rescaled to a fixed number of charge carriers (1200 electron hole pairs). For cloud sizes with more than 1.2$\times$10$^5$ electron hole pairs pulse distortions due to the charge carrier density are visible. The right graph shows pulse shapes for 1200 electron hole pairs starting an the junction (high field side) and the influence of the order of the time integration scheme. For comparison the laser pulse structure is also shown. The applied voltage for both graphs is 80~V.\label{fig-diode-small-charge-limit}}
\end{figure*}

Along the main drift path of the charge cloud the grid is refined (central part of the cylinder). The illuminated side of the cylinder (top or bottom) is refined as well to resolve the small initial cloud. The distribution of created charge carriers is parameterized as
\begin{eqnarray}
 N(x,y,z,t)&=&C_0 L(t) exp\left( f(x,y,z)\right) \\
 f(x,y,z) &=&-(x/2\sigma_x)^2-(y/2\sigma_y)^2-|z-z_s|/\lambda_{abs}
\end{eqnarray}
with $\sigma_x$~=~$\sigma_y$~=~10~$\upmu$m, $z_s$ top or bottom, $\lambda_{abs}$~=~3~$\upmu$m the absorption length of the laser light, $L(t)$ the laser pulse shape and $C_0$ a constant to ensure that the correct number of electron hole pairs is created. The laser pulses shape is shown in the right graph of Figure~\ref{fig-diode-small-charge-limit}, its peak is very narrow (FWHM~$<$~100~ps) and its total length is approximately 1~ns.


The p$^+$ doping profile is the result of a one dimensional
process simulation (compare Figure~\ref{fig-diode-gridA1}). It is of short range compared to the n$^+$ doping profile. 

The n$^+$ doping profile was determined by spreading resistance measurements. 
The bulk doping concentration of $8.2\times10^{11}{\rm cm^{-3}}$ is marked by the dotted horizontal line in the right graph of Figure~\ref{fig-diode-gridA1}.

\subsection{Employed mobility parameters}

All results presented here are based on charge carrier mobilities reported in literature \cite{selberherr84} (summarized in Table \ref{pars}), but using $\beta_n=1$ (instead of $\beta_n=2$) for the field dependent mobility reduction (labeled literature mobility), unless mentioned otherwise. For comparison two other models have been used as well. The second model is the same as mentioned above but without field dependent mobility reduction (labeled constant mobility). The third model uses the same models for lattice, ionized and unionized impurity scattering as well as the same model for carrier-carrier scattering, but the field dependent parameters for $<$100$>$ crystal orientation described in \cite{becker2} were used (labeled fitted mobility, summarized in Table~\ref{mobpar}).

\section{Simulation results}
\subsection{Test of the numerical methods} 

To test the numerical methods and the models including their parameters a set of simulations with low numbers of charge carriers was done. 
The simulated currents were rescaled to a fixed charge after the simulation process for comparison and are shown in the left graph of Figure~\ref{fig-diode-small-charge-limit}. 

\begin{figure*}[t!]
  $\begin{array}{cc}
  \includegraphics[width=0.5\textwidth]{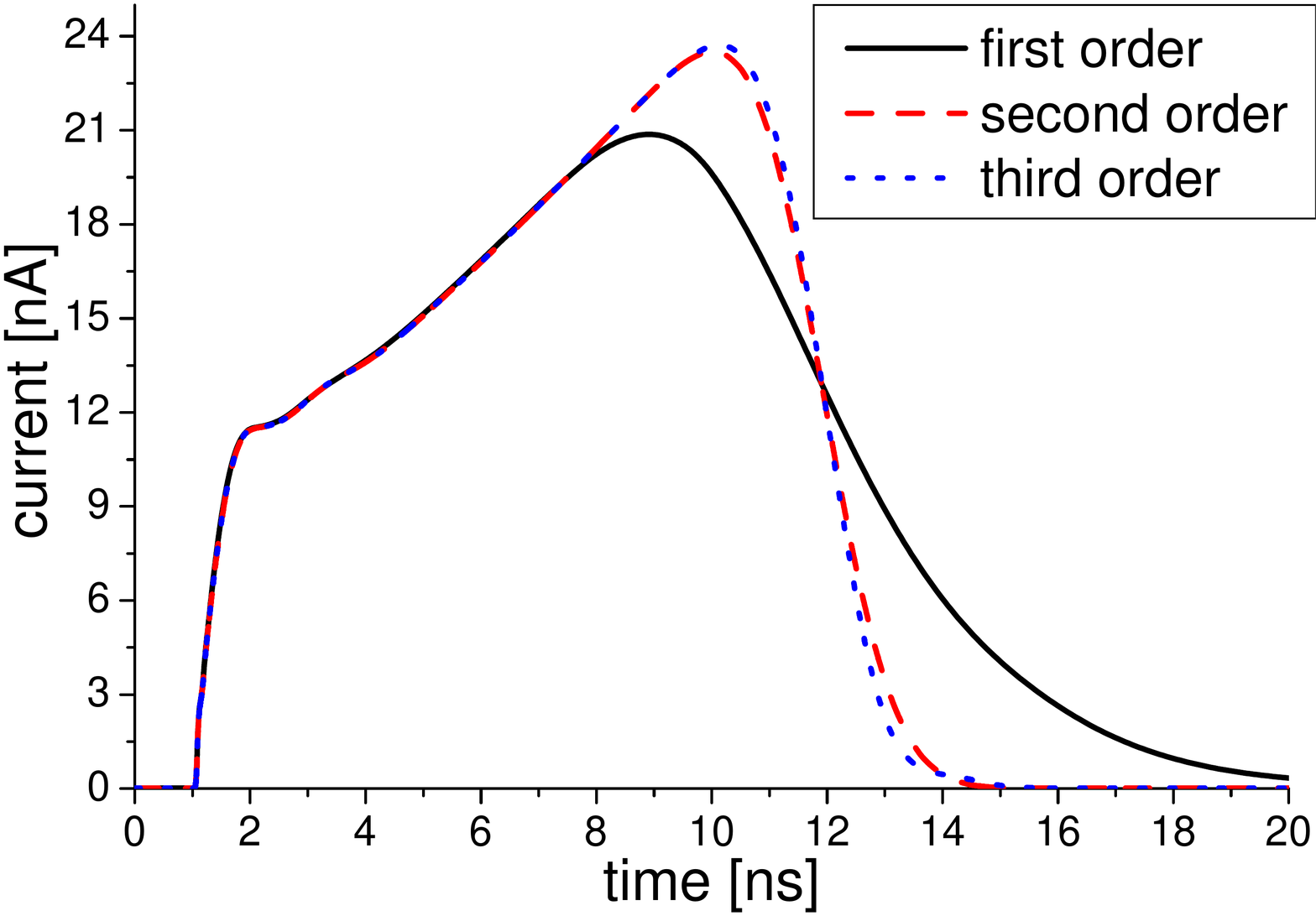} &
 	\includegraphics[width=0.5\textwidth]{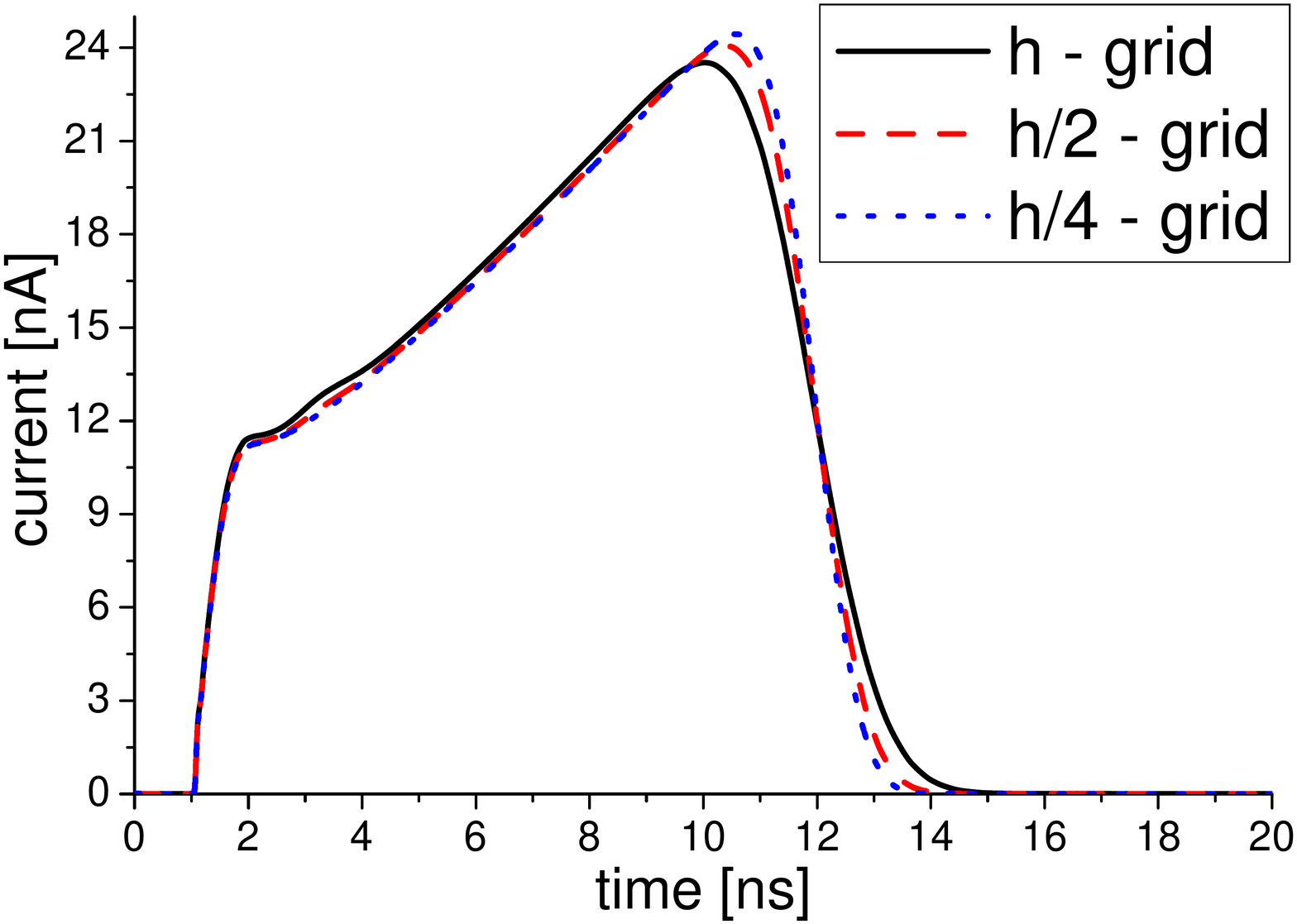} 
\end{array}$
  \centering
  \caption{Comparison of typical pulses of charge carriers created opposite to the junction with different orders in the time integration scheme (left) and different grid sizes (right). Both graphs show pulses for 1200 electron hole pairs starting opposite to the junction for an applied bias of 80~V.\label{fig-diode-time-space-err-low}}
\end{figure*}

Typical discretization errors are evaluated by repeating computations for one number of charge carriers with different spatial and time step sizes and a different time integration order. Effects of different time integration orders for junction side creation of charge carriers is shown in the right graph of Figure~\ref{fig-diode-small-charge-limit}. Effects for charge carrier creation opposite to the junction are shown in Figure~\ref{fig-diode-time-space-err-low}. The left graph of Figure~\ref{fig-diode-time-space-err-low} shows the effects of different time integration schemes, the right graph shows the effect of different grid sizes.

As expected the implicit Euler scheme (i.e. the first order time integration scheme, see appendix for details) results in too much energy dissipation. This can 
be observed as additional diffusion and is seen in the pulse shape of the currents shown in the right graph of Figure~\ref{fig-diode-small-charge-limit} and the left graph of Figure~\ref{fig-diode-time-space-err-low}. 

For the implicit Euler scheme the current starts to decrease earlier due to the artificially increased diffusion and shows a long tail.

\begin{figure*}[t!]
  $\begin{array}{cc}
  \includegraphics[width=0.5\textwidth]{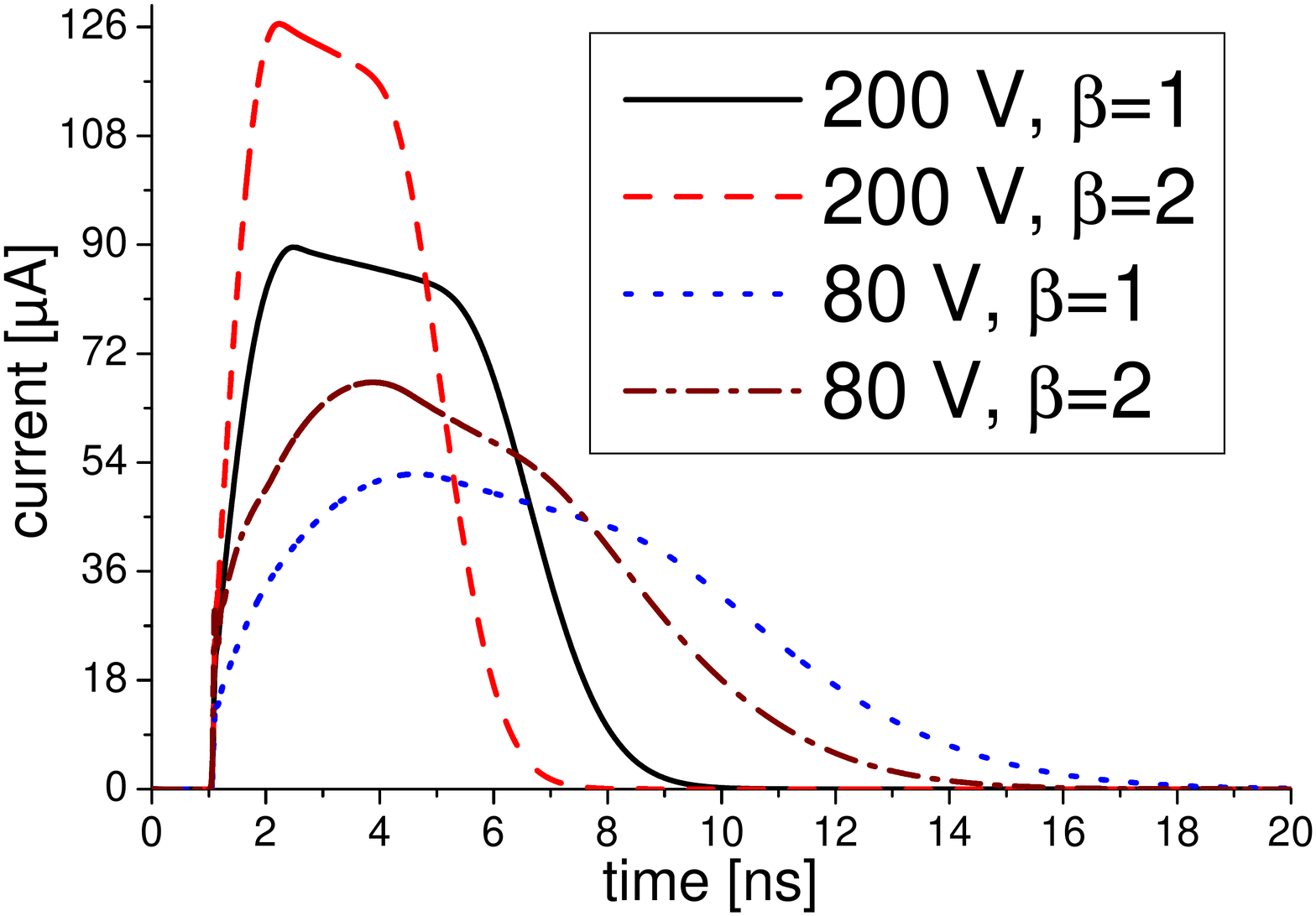} &
 	\includegraphics[width=0.5\textwidth]{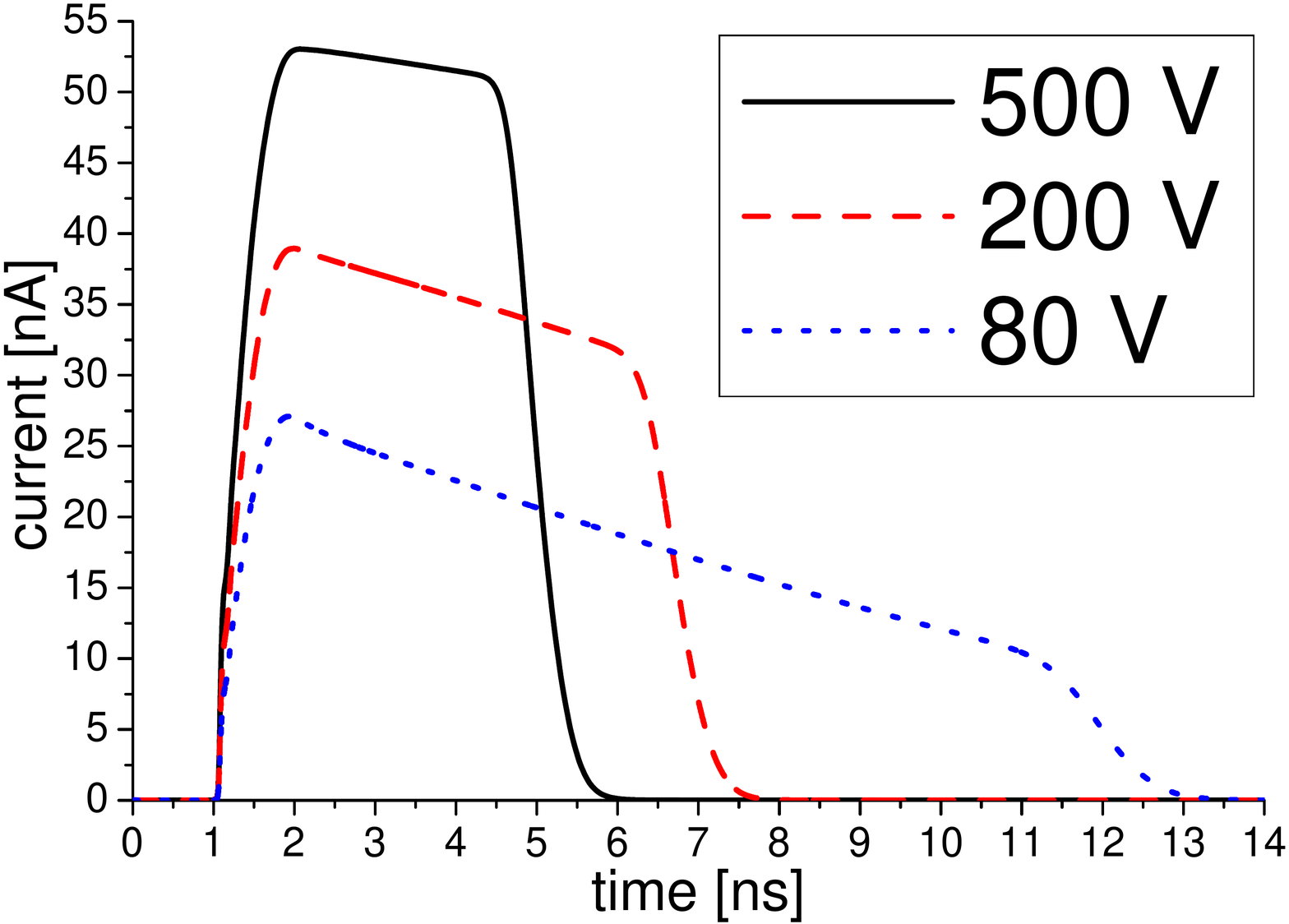} 
\end{array}$
  \centering
  \caption{The left graph shows the influence of $\beta_n$ on pulses for 2.9$\times$10$^6$ electron hole pairs for 80~V and 200~V applied voltage. The right graph shows pulse shapes for 1200 electron hole pairs and $\beta_n=1$ as function of the applied voltage. For both graphs the electron hole pairs were created at the junction side.\label{fig-diode-beta}}
\end{figure*}

The differences between second and third order time integration schemes are small 
compared to the difference between the implicit Euler and the higher order schemes.

As a consequence all simulations which were compared to measurements, do not use the first order scheme in the calculations.

The spatial discretization errors are checked by using the $h$, $h/2$, and $h/4$ grids explained above. 
Compared with the time discretization error of the Euler scheme the space
discretization error shown in the right graph of Figure~\ref{fig-diode-time-space-err-low} is small.

The influence of spatial discretization errors is considered negligible.

\subsection{Interpretation of the results} 

When the charge cloud is created at the junction (i.e. high field) side the rise time of the current measures the cloud creation and charge carrier separation time.
For small clouds the current peak 
corresponds to the situation where all charge carriers are drifting. In this case all holes are quickly collected at the electrode and the current is due to drifting electrons.

The almost linear reduction of the current is due to the lower average velocity at lower electric fields.

Around the 'knee' the first charge carriers reach the contact and the total number of charge carriers in the sensor drops. 
The strong decrease of the current represents the removal of charge carriers from the bulk volume at the contact and thus reflects the shape of the charge cloud along the drift direction. 
This shape is very sensitive to the enhanced diffusion of the implicit Euler scheme.

When the charge cloud is created opposite to the junction (i.e. low field) side the current shows a steep increase while the cloud it created and separates. In this case all electrons are quickly collected at the electrode and the current is due to drifting holes.

The almost linear increase of the current is due to the higher drift velocity at higher electric fields. 

Around the peak the charge cloud reaches the contact and the number of charge carriers in the bulk material drops. 
Like in the case of illumination from the junction side the decrease of the current represents the removal of charge carriers from the sensor volume at the contact and its shape is very sensitive to the enhanced diffusion of the implicit Euler scheme.

\begin{figure*}[t!]
 	\includegraphics[width=1.0\textwidth]{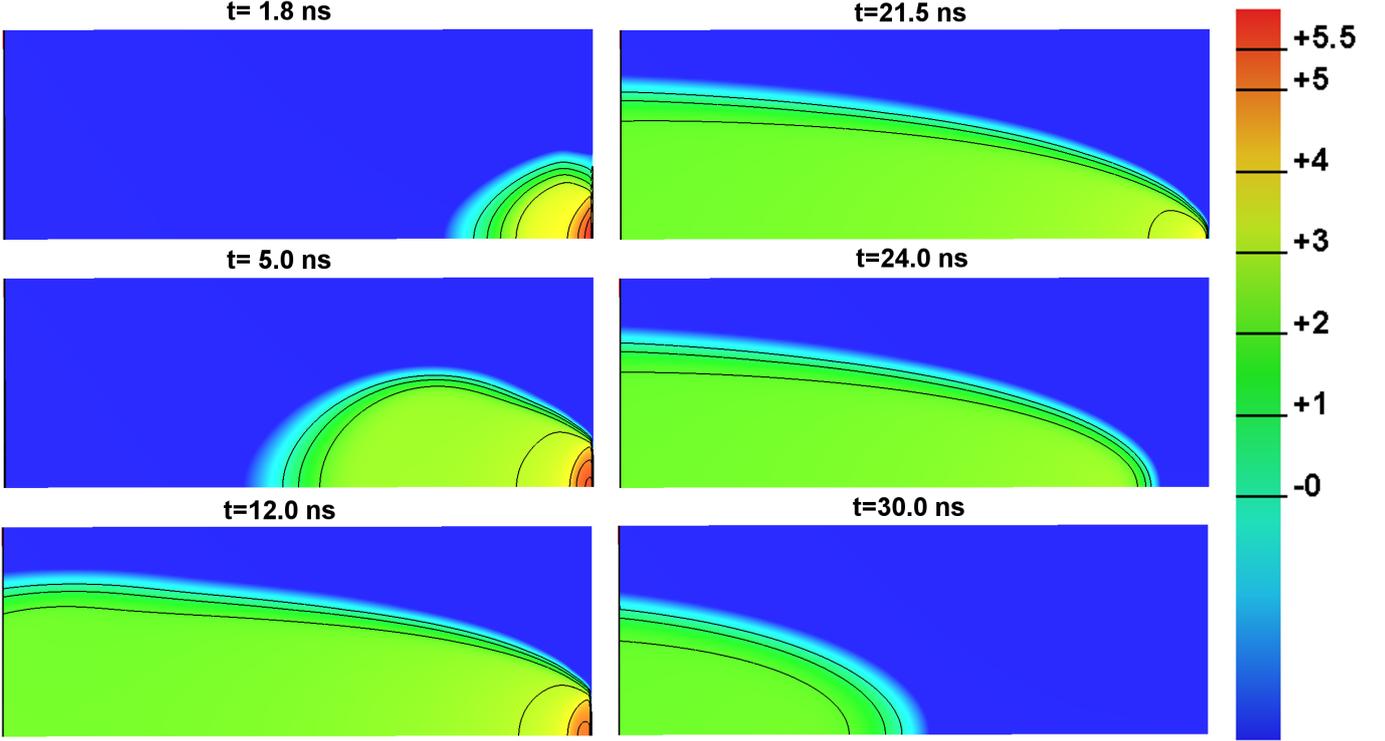} 
  \centering
  \caption{Evolution of a plasma cloud in space and time. The logarithm to the base of ten of the hole density divided by 10$^{10}$~cm$^{-3}$ is color-coded for a cut along the y axis. The simulation volume is 100~$\upmu$m high (top to bottom) and 280~$\upmu$m wide (left to right). 11$\times$10$^6$ electrons and holes are created on the right side (opposite to the junction) and holes drift to the left. A bias of 200~V is applied.\label{plasma}}
\end{figure*}

At the applied voltages neither the drift velocity of electrons nor the drift velocity of holes is saturated. Thus a charge cloud of low density traveling in the direction of a decreasing electrostatic field is compressed, while a low density charge cloud traveling in the other direction is elongated. 

The mobility reduction due to large fields has a significant influence on 
the pulse shape. 
A strong influence of $\beta_{n,p}$ is observed as the drift velocity in the entire volume is neither in the ohmic regime nor saturated. 
Figure~\ref{fig-diode-beta} illustrates the influence of the 
model parameter $\beta_n=1$ or $\beta_n=2$. $\beta_n=1$ results in more pronounced spatial fronts and $\beta_n=2$ causes a lower velocity reduction and thus 
smaller flight times at high voltages. 

While simulations with $\beta_n=2$ show significantly different pulses than observed in measurements, simulations with $\beta_n=1$ produce pulses which are very similar to the measurements, as shown in Figure~\ref{small_electrons} for junction side illumination for defocused laser light.

\subsection{Time evolution of the spatial distribution of charge carriers}

Figure \ref{plasma} and the movies in the online version of this article show simulations of the time evolution of the hole density for 11$\times$10$^6$ electron hole pairs and a bias of 200~V. The electron hole pairs are created by laser light of 660~nm light focused to 10~$\upmu$m, injected opposite to the p$^+$n junction.

From the drift-diffusion calculations it is concluded, that the plasma cloud does not expand, instead charge carriers are continuously released from the plasma region and thus form a conductive channel connecting both electrodes.

\section{Comparison of simulations and measurements}

\begin{figure}[tb!]
  \includegraphics[width=0.5\textwidth]{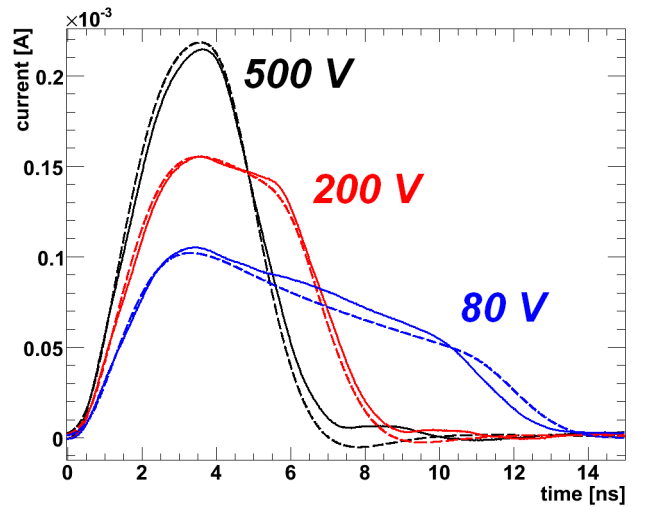}
  \centering
  \caption{Results for junction side illumination using defocused laser light to avoid plasma effects. Simulations with literature mobility are shown as dashed lines and measurements as solid lines. As simulations and measurements are very similar the use of $\beta_n=1$ is justified.\label{small_electrons}}
\end{figure}

Electron hole pairs have been created with 660~nm laser light of different intensities.
The measurements for low intensities allow to verify the simulations, especially the field dependence of the mobility.
For high intensity illuminations plasma effects dominate experimental results and simulations.

\begin{figure*}[tb!]
  \mbox{\bf 1$\times$10$^6$ electron hole pairs} \\
  \includegraphics[totalheight=0.25\textheight,width=0.8\textwidth]{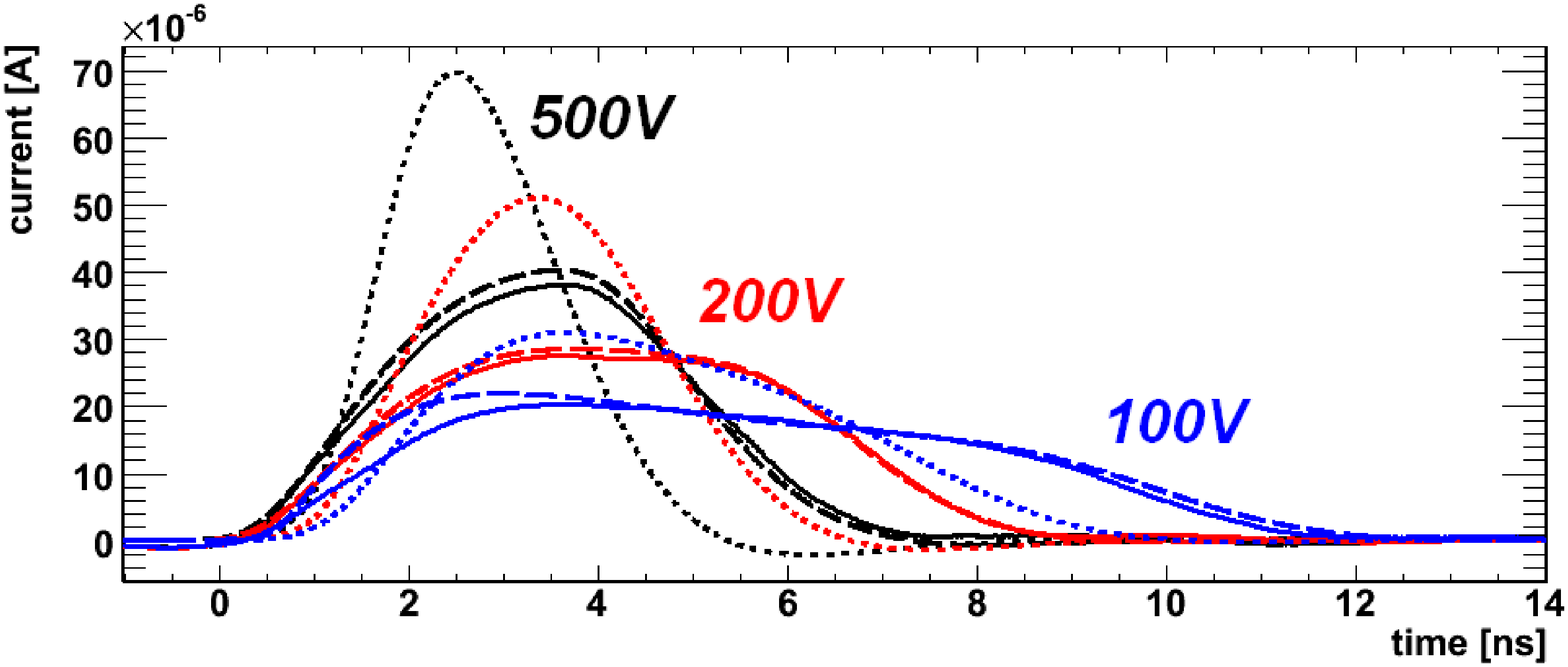} 
  \mbox{\bf 10$\times$10$^6$ electron hole pairs} \\
 	\includegraphics[totalheight=0.25\textheight,width=0.8\textwidth]{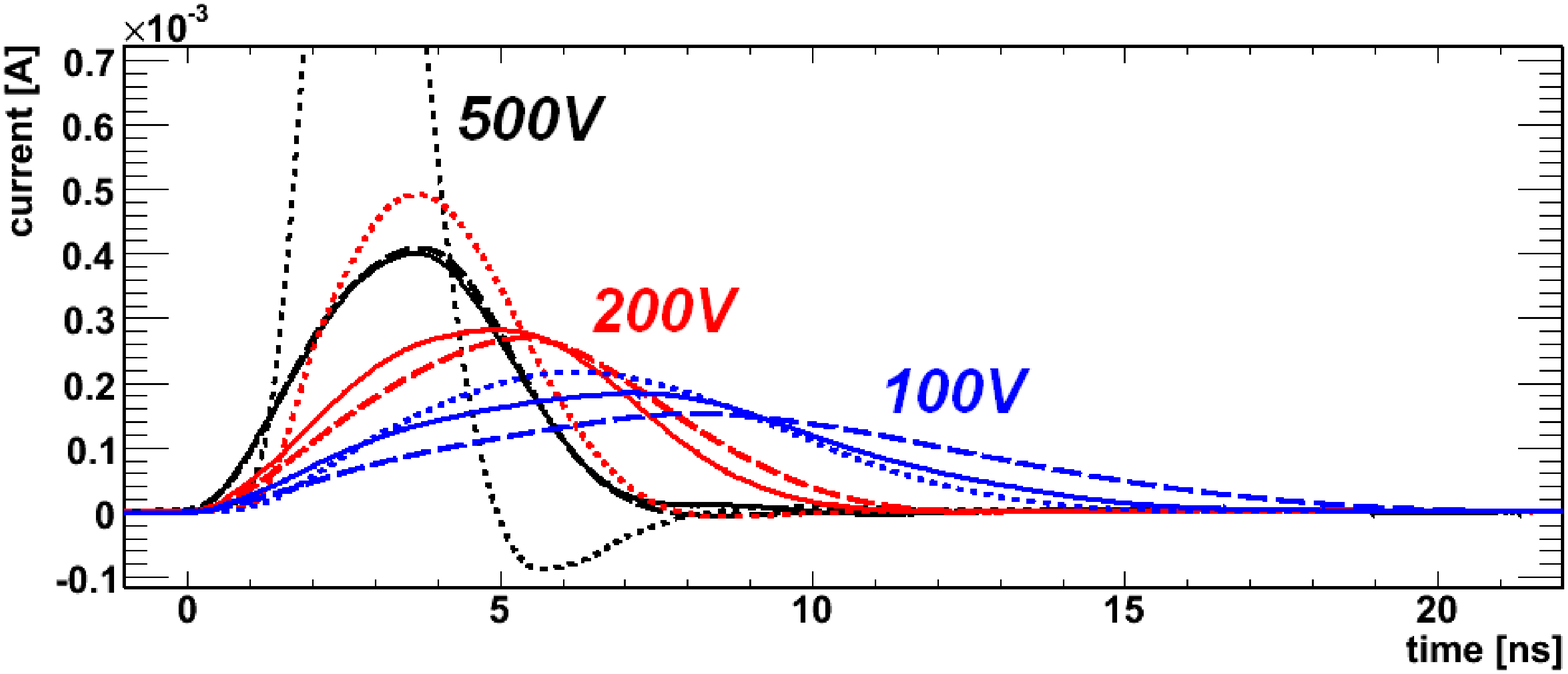} 
  \mbox{\bf 97$\times$10$^6$ electron hole pairs} \\
  \includegraphics[totalheight=0.25\textheight,width=0.8\textwidth]{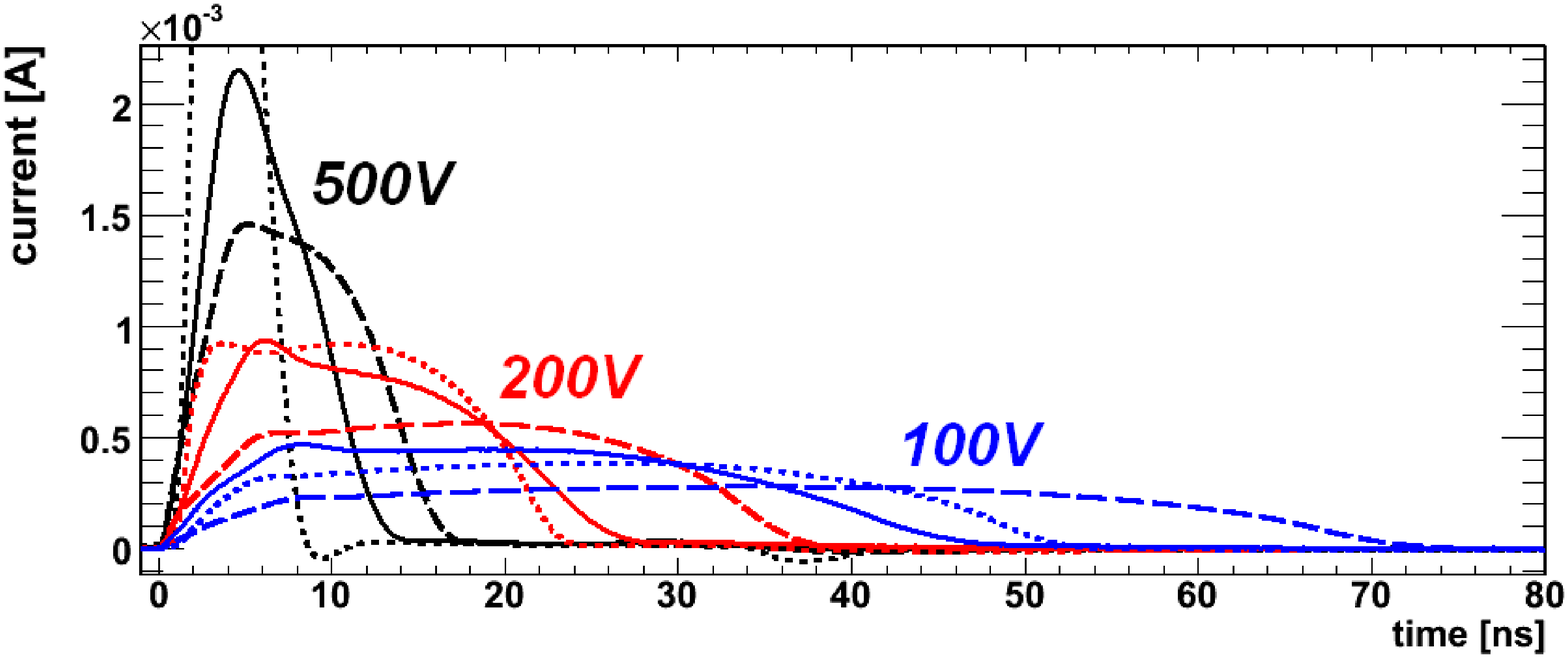}
  \centering
  \caption{Measurements for junction side illumination using focused laser light (solid lines) compared to simulations. Simulations with literature mobility are shown as dashed lines, simulations with constant mobility as dotted lines. While acceptable agreement can be observed for low densities, high densities show a shorter pulse duration than the simulations. Note the different scales on the time axes.\label{fig-electrons1}}
\end{figure*}

The observed current pulses show small ringing features (oscillations), due to the presence of inductances in the readout circuit. If the gradient on a decreasing slope is large enough the associated ringing may produce and undershoot into negative currents. This behavior is modeled by the equivalent circuit shown in Figure \ref{circuit}. The undershoot is neither expected nor found in the simulations before convolution with the transfer function (Figure \ref{transfer}).

In the simulations peak densities of $ n p/n_i^2 \approx 10^{10}$ are observed, at this density carrier-carrier scattering is still of small influence. Simulations without mobility reduction due to carrier-carrier scattering showed the same pulse durations and are not presented here. Artificially enhanced carrier-carrier scattering increases the pulse duration as expected, while the measurements show shorter pulses than the simulations.

\subsection{Junction side illumination}

Injection on this side allows to check the transport properties of electrons separately from those of holes, as
holes reach the close by electrode quickly.

Increasing the number of created carriers increases the plasma effects. Inspection of the simulations
shows peak values of $n p /n_i^2 \approx 10^{10}$ in the plasma cloud for the case 
of 97$\times$10$^6$ generated charge carriers.

Figures \ref{small_electrons} and \ref{fig-electrons1} show the comparison of simulated and measured currents for junction side illumination.

Simulations and measurements for low density clouds are very similar and
acceptable agreement between simulation and measurement is observed for 1$\times$10$^6$ electron hole pairs.  Thus it can be concluded that the low density transport properties of electrons are reasonably well modeled. At higher charge carrier densities the current pulse shapes deviate from the low density case due to the lifetime of the plasma cloud. The current is determined by the release of charge carriers from the plasma cloud. 
The deviation between measurements and simulations increases with increasing plasma density.

As seen in Figure~\ref{fig-electrons1}, the current rises slowly at the beginning of
the pulse for 97$\times$10$^6$ electron hole pairs in simulation and measurement.
The rise is followed by an approximately constant current, until the final decrease shows a similar time constant
as in the 10$\times$10$^6$ electron hole pairs case.

In the simulations the peak at the beginning is due to the removal
of the low density periphery of the cloud. The remaining high density core
has an ellipsoidal shape and shrinks with time as well as the maximum density in the plasma (see Figure \ref{plasma}). The barycenter of the plasma slowly moves away from the junction.

Using a constant mobility speeds up the release of charge carriers from the plasma and their drift in the rest of the sensor volume. Thus pulses calculated using a constant mobility are systematically too short, except for the pulse obtained for 97$\times$10$^6$ electron hole pairs at 100~V bias, which is too long by 4\% of the pulse length.

\subsection{Illumination opposite to the junction}

\begin{figure}[tb!]
  \includegraphics[width=0.5\textwidth]{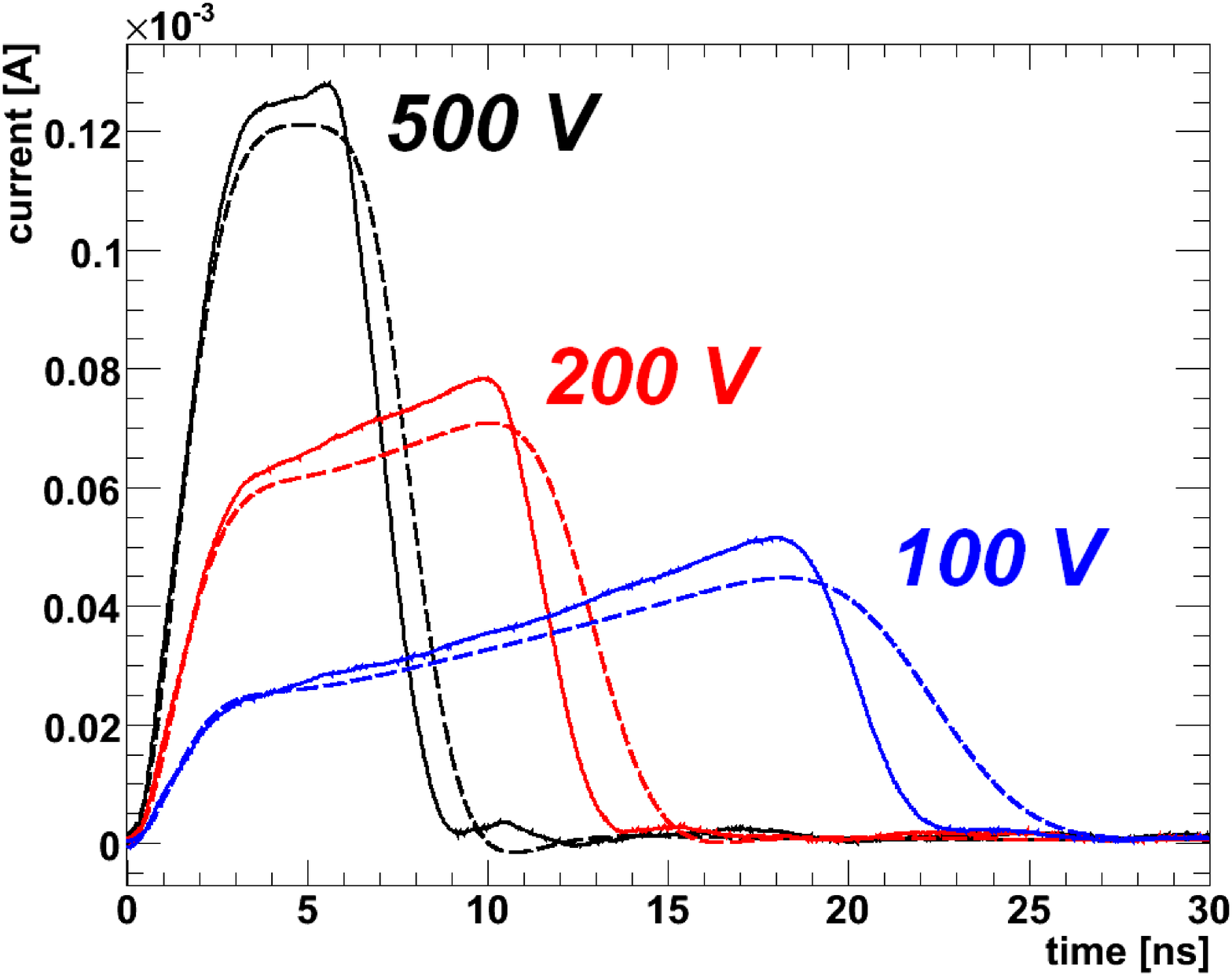}
  \centering
  \caption{Results for illumination opposite to the junction using defocused laser light to avoid plasma effects. Simulations with literature mobility are shown as dashed lines and measurements as solid lines. \label{small_holes}}
\end{figure}

\begin{figure*}[tb!]
  \mbox{\bf 1$\times$10$^6$ electron hole pairs}\\
  \includegraphics[totalheight=0.25\textheight,width=0.8\textwidth]{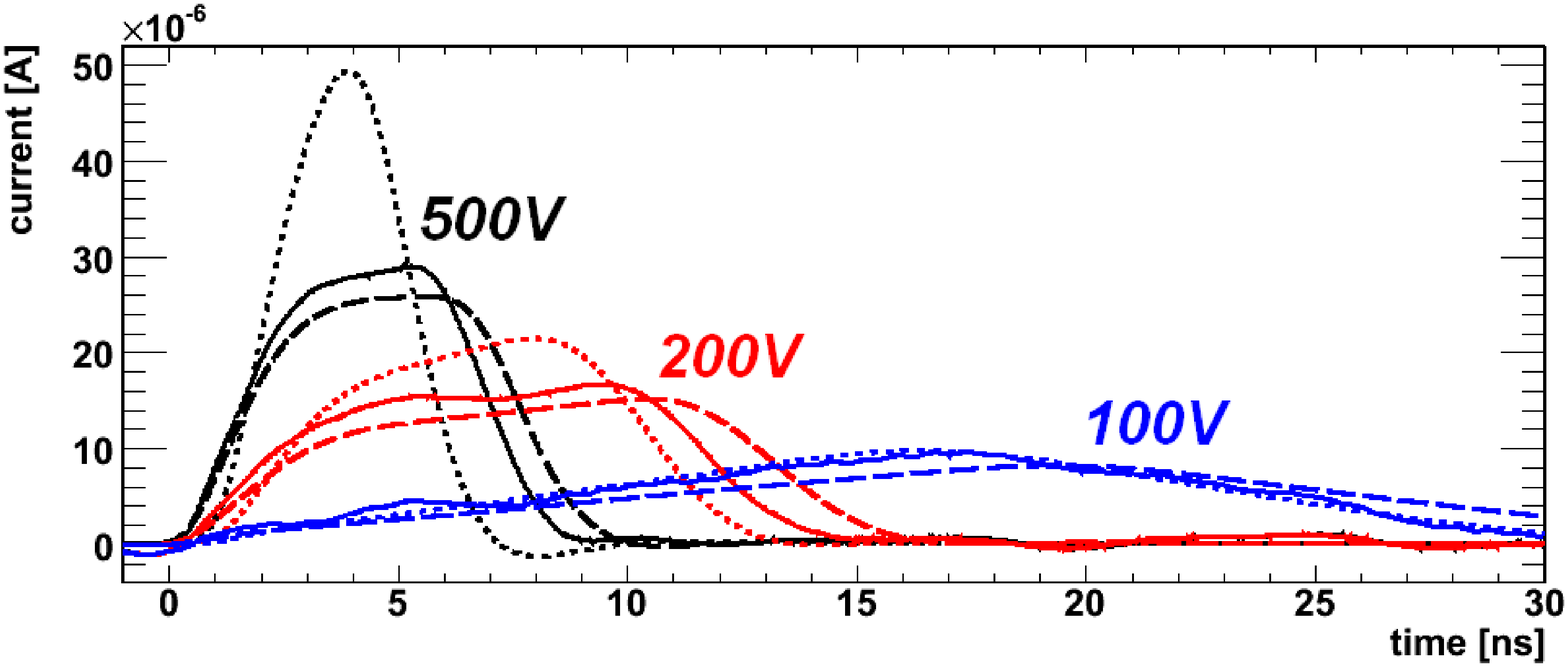} 
  \mbox{\bf 11$\times$10$^6$ electron hole pairs}\\
 	\includegraphics[totalheight=0.25\textheight,width=0.8\textwidth]{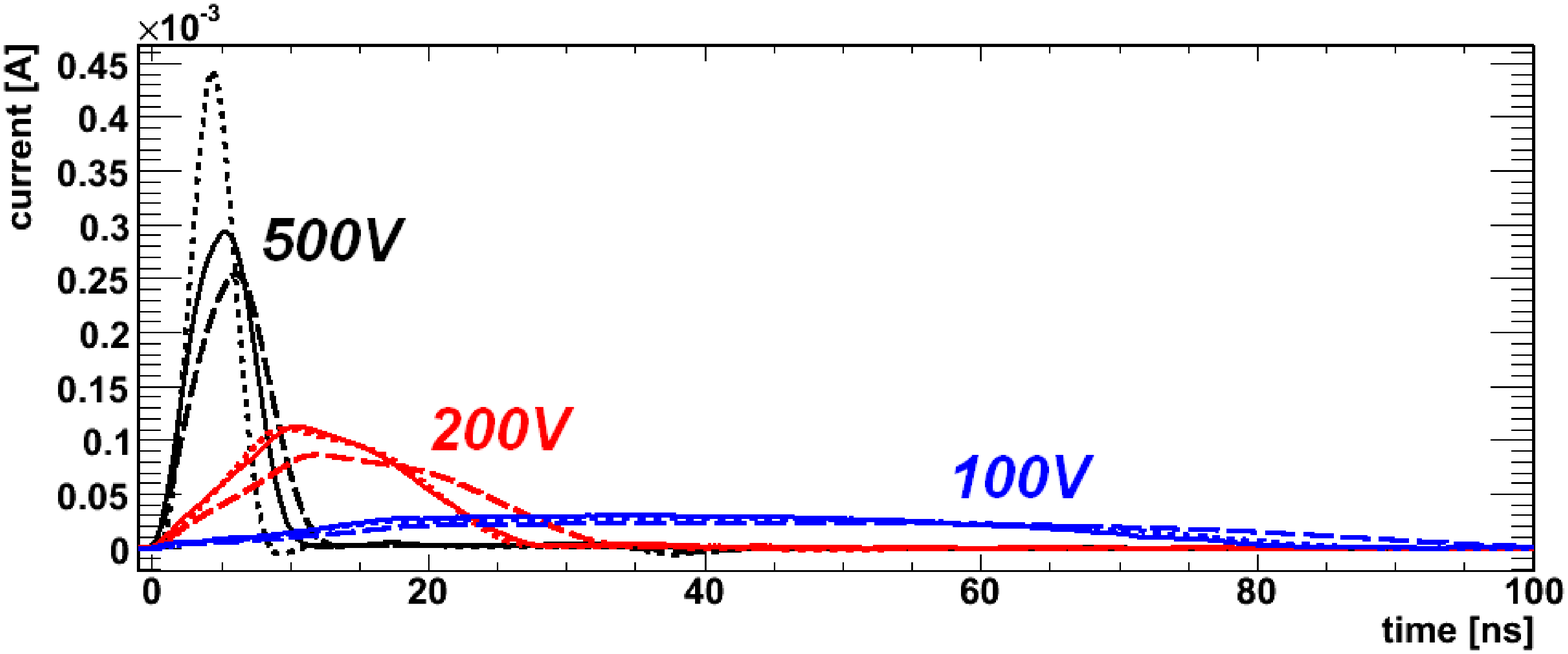} 
  \mbox{\bf 103$\times$10$^6$ electron hole pairs}\\
  \includegraphics[totalheight=0.25\textheight,width=0.8\textwidth]{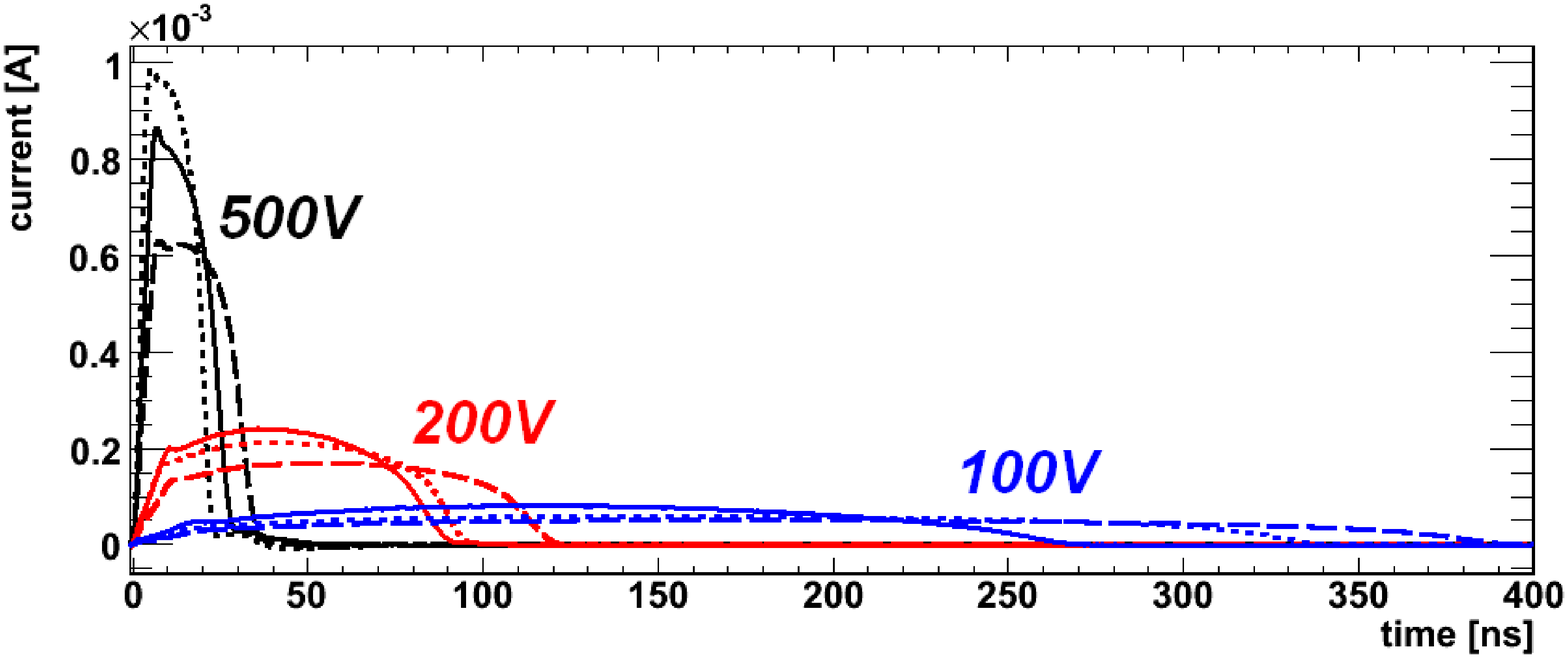}
  \centering
  \caption{Measurements for injection opposite to the junction using focused laser light (solid lines) compared to simulations. Simulations with literature mobility are shown as dashed lines, simulations with constant mobility as dotted lines. Measurements systematically show a shorter pulse duration than the simulations. Note the different scales on the time axes.\label{fig-holes1}}
\end{figure*}

This situation allows to study the transport properties of the holes. Contrary to electrons, holes move towards the high field region (junction).
Figures \ref{small_holes} and \ref{fig-holes1} show the comparison of simulated and measured currents for this case.

A qualitative agreement between simulation and measurement is observed for the measurement with defocused laser light and 1$\times$10$^6$ electron hole pairs. However the simulated pulses are systematically too long. 

For 11$\times$10$^6$ electron hole pairs the 100~V and 200~V curves show clear deviations, while at 500~V the deviations are at the level observed for 1$\times$10$^6$ electron hole pairs. The transient behavior between the low density regime and first plasma effects observed for junction side illumination can also be observed in this case.

When 103$\times$10$^6$ electron hole pairs were created, both the simulated and the measured current pulse become very long compared to the current pulses observed for low densities. While qualitatively the simulations produce pulses with similar durations, the simulated pulses are systematically too long.

As in the case of illumination on the junction side, using a constant mobility speeds up the release of charge carriers from the plasma and their drift in the rest of the sensor volume. Thus pulses calculated using a constant mobility are systematically too short as well, except for the pulse obtained for 11$\times$10$^6$ electron hole pairs at 100~V bias (2.4\% too long) and 103$\times$10$^6$ electron hole pairs at 200~V (5.6\% too long) and 100~V bias (25.9\% too long).

\section{Discussion and Conclusions}

\begin{table*}[t!]
\centering
\begin{tabular}{c|c|c|c|c}
Illumination & $N_{e,h}$ [10$^6$] & $U_{bias}$ [V]& duration [ns] & difference \\
\hline
junction side 	 & -   & 500 & $\emptyset$/7.0/6.8    &$\emptyset$ / -2.9\% \\
junction side 	 & -   & 200 & $\emptyset$/8.5/8.4    &$\emptyset$ / -1.2\% \\
junction side 	 & -   & 80  & $\emptyset$/13.0/13.5  &$\emptyset$ / 3.8\% \\
opp. to junction & -   & 500 & $\emptyset$/9.0/10.0   &$\emptyset$ / 11.1\% \\
opp. to junction & -   & 200 & $\emptyset$/13.5/15.0  &$\emptyset$ / 11.1\% \\
opp. to junction & -   & 100 & $\emptyset$/23.0/27.0  &$\emptyset$ / 17.4\% \\
\hline
junction side 	 & 1   & 500 & 5.0/7.0/7.0   &-28.6\% / 0.0\% \\
junction side 	 & 1   & 200 & 6.5/9.0/9.0   &-27.8\% / 0.0\% \\
junction side 	 & 1   & 100 & 10.0/12.0/12.3 &-16.7\% / 2.5\% \\
junction side 	 & 10  & 500 & 5.0/7.0/7.0   &-28.6\% / 0.0\% \\
junction side 	 & 10  & 200 & 8.0/11.0/12.0 &-27.3\% / 9.1\% \\
junction side 	 & 10  & 100 & 15.0/17.0/20.0 &-11.8\% / 17.6\% \\
junction side 	 & 97  & 500 & 8/14/18     &-42.9\% / 28.6\% \\
junction side 	 & 97  & 200 & 24/28/38     &-14.3\% / 35.7\% \\
junction side 	 & 97  & 100 & 52/50/75     &4.0\% / 50.0\% \\
opp. to junction & 1   & 500 & 7.0/9.0/10.0  &-22.2\% / 11.1\% \\
opp. to junction & 1   & 200 & 13/15/17     &-13.3\% / 13.3\% \\
opp. to junction & 1   & 100 & 31/31/38     &0.0\% / 22.6\% \\
opp. to junction & 11  & 500 & 9/12/14     &-25\% / 16.6\% \\
opp. to junction & 11  & 200 & 27/28/35     &-3.6\% / 25.0\% \\
opp. to junction & 11  & 100 & 87/85/100    &2.4\% / 17.6\% \\
opp. to junction & 103 & 500 & 25/30/35     &-16.7\% / 16.7\% \\
opp. to junction & 103 & 200 & 95/90/120    &5.6\% / 33.3\% \\
opp. to junction & 103 & 100 & 340/270/390   &25.9\% / 44.4\% \\
\end{tabular}
\caption{Pulse durations for all measurements ($t_{meas}$) presented in this work compared to their simulations with the literature mobility model ($t_{lit}$) and the constant mobility model ($t_{const}$). The column labeled $N_{e,h}$ lists the number of created electron hole pairs; a '-' indicates the low density measurement and simulation. The column labeled $U_{bias}$ lists the applied bias voltage. The column labeled duration lists $t_{const}/t_{meas}/t_{lit}$. The column labeled difference lists ${(t_{const}-t_{meas})}/{t_{meas}}$ and ${(t_{lit}-t_{meas})}/{t_{meas}}$. A '$\emptyset$' indicates that no corresponding simulation was performed. The measurements have been performed with 660~nm light focused to a Gaussian spot with $\sigma_{laser}$~=~3~$\upmu$m on a p$^+$nn$^+$ diode with a thickness of 280~$\upmu$m and an effective doping of 8.2$\times$10$^{11}$~cm$^{-3}$.}
\label{minmax}
\end{table*}



At low charge carrier densities the simulations for the collection of electrons (junction side illumination) agree well with the measurements. This gives us confidence in the simulation as well as in the model used for the electron transport. For the collection of holes (opposite side illumination), the simulated pulses are typically 10\% longer than the measured ones. This may indicate a problem with the simulation of the charge cloud separation at low fields or with the model used for the hole transport. In \cite{becker2} the experimental data has been used to determine the parameters of electron and hole transport. 

High charge carrier densities result in significant distortions of the pulse shapes, both in data as well as in simulations. For the highest charge densities investigated ($\approx$~10$^8$ electron hole pairs) the measured pulse lengths increase by up to a factor 5 for electrons and a factor 10 for holes. The results are charge collection times up to 250~ns for a detector operated well above depletion! The simulations qualitatively describe the data and provide an understanding of the pulse shape for high charge densities: An initial rise due to the collection of charges from the periphery of the charge cloud is followed by an approximately constant current due to the release of charges from a shrinking plasma cloud. 


A number of attempts have been made to understand the discrepancies. 

\begin{figure*}[tb!]
  \includegraphics[width=0.7\textwidth]{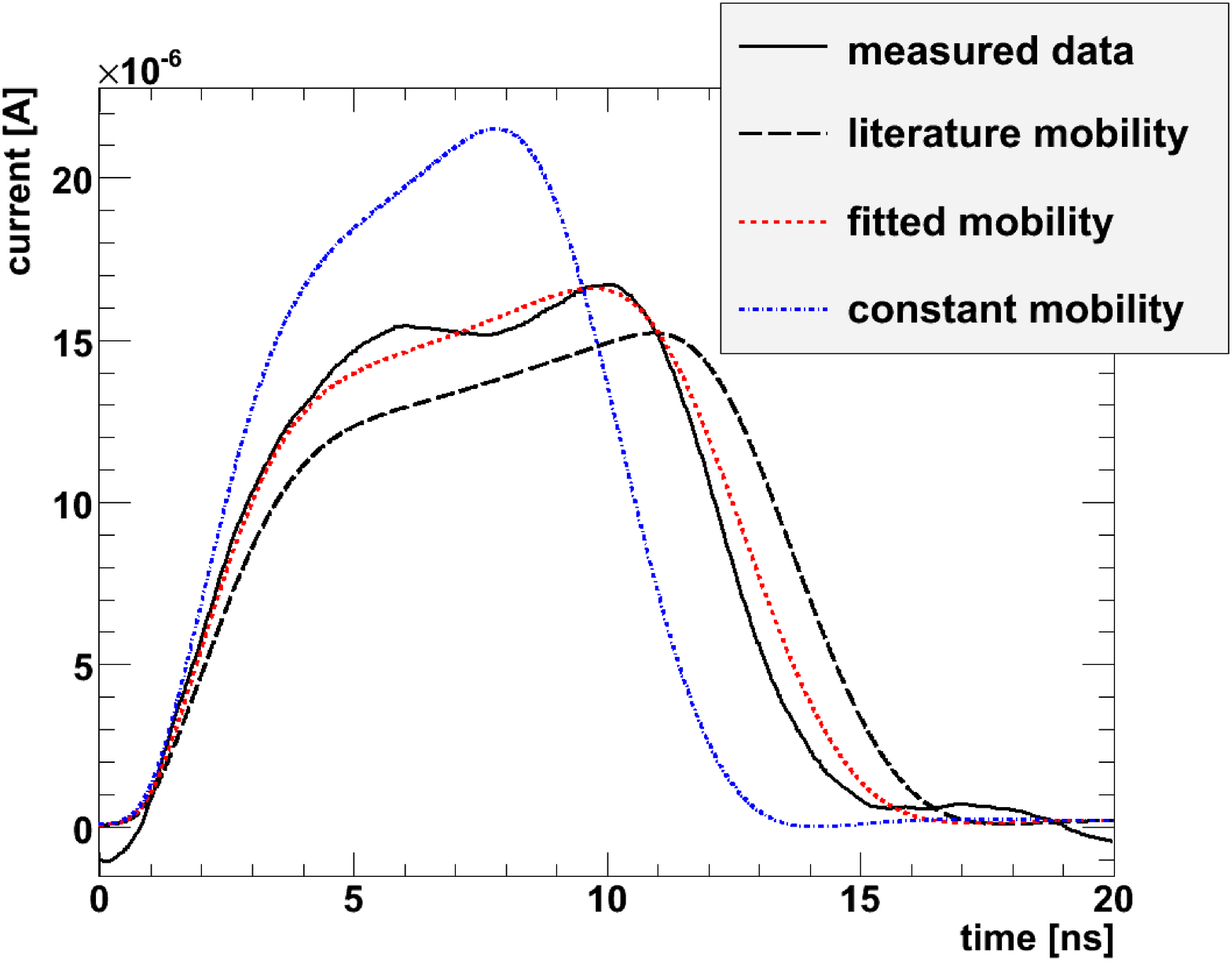}
  \includegraphics[width=0.7\textwidth]{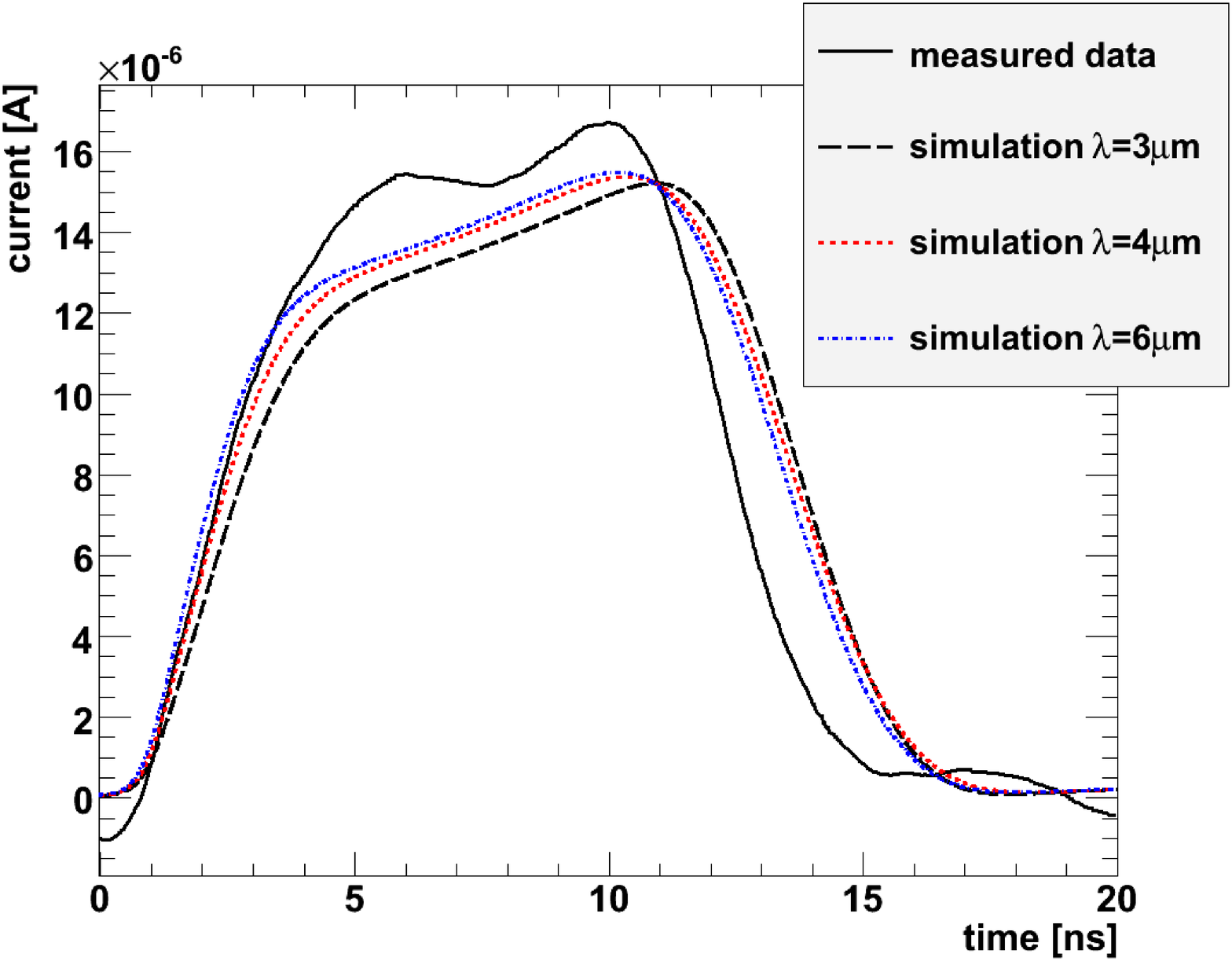}
 \centering
  \caption{Effects of different parameters on the simulated current pulses for 1$\times$10$^6$ electron hole pairs at 200~V applied voltage. The upper graph shows the influence of different mobility models. The lower graph shows the influence of different distributions of the initial charge cloud. Although lower density clouds dissolve faster, the effects of different attenuation lengths are small compared to the effects of a constant mobility model.\label{mod_pars_1}}
\end{figure*}

\begin{figure*}[tb!]
  \includegraphics[width=0.7\textwidth]{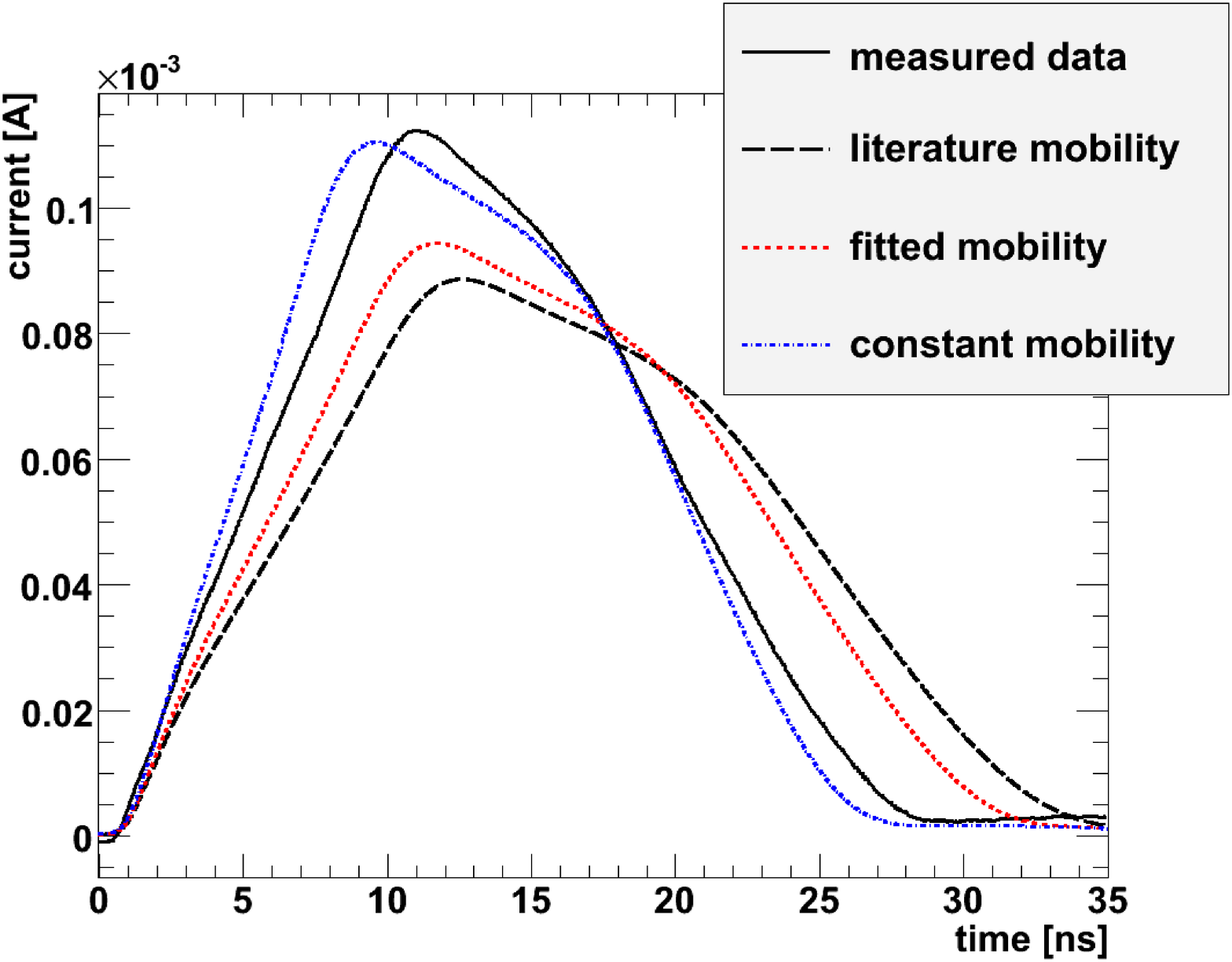}
  \includegraphics[width=0.7\textwidth]{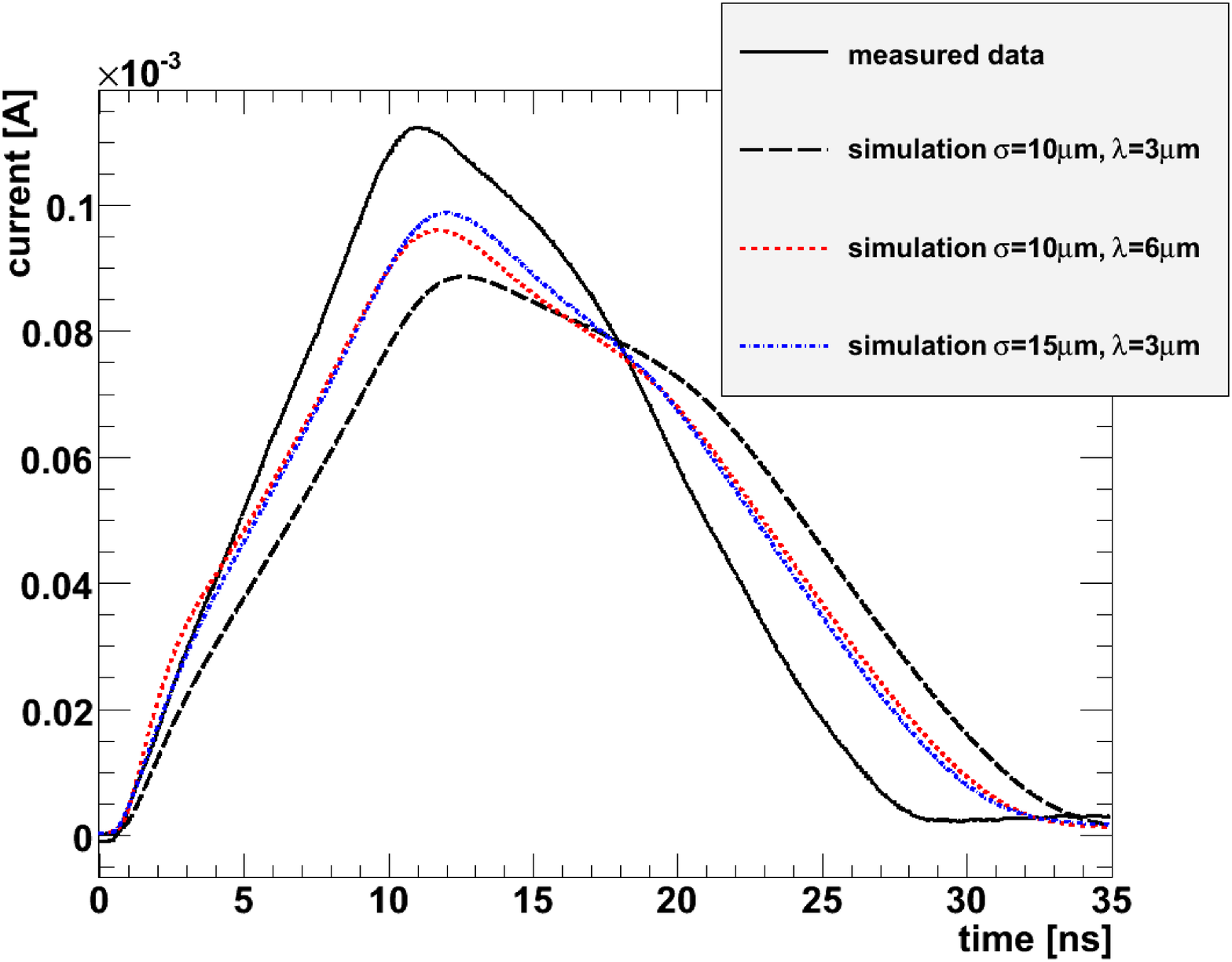}
 \centering
  \caption{Effects of different parameters on the simulated current pulses for 11$\times$10$^6$ electron hole pairs at 200~V applied voltage. The upper graph shows the influence of different mobility models. The lower graph shows the influence of different distributions of the initial charge cloud. Although lower density clouds dissolve faster, the effects of different initial distributions are small compared to the effects of a constant mobility model.\label{mod_pars_11}}
\end{figure*}

To estimate the effect of having an initial charge carrier distribution which is different from the light profile, simulations with different widths of the initial charge carrier distribution and absorption lengths have been done for 1$\times$10$^6$ and 11$\times$10$^6$ electron hole pairs created opposite to the junction. The results for 200~V applied bias are presented in the upper graphs of Figure~\ref{mod_pars_1} and \ref{mod_pars_11}, showing that an increase in width by 50~\% or an increase in absorption length by 100~\% is not sufficient to produce a pulse which is as short as the measurement. 

Using literature data of the mobility, the low density limit for holes (Figure~\ref{small_holes}) could not be reproduced, indicating the need for changes in the mobility parameterizations. Using the fitted mobility parameterization for holes the current pulse for 1$\times$10$^6$ electron hole pairs could be well reproduced (Figure~\ref{mod_pars_1}, bottom).

Although certainly not realistic, we have made simulations assuming that the mobility is independent of the electric field (constant mobility model). The pulses obtained are systematically shorter than the measured ones (except for the highest intensity at low bias voltage). However the mobility model from the literature and the constant mobility model give the interval in which the measured pulse durations are found. Table~\ref{minmax} summarizes the results. The results assuming the constant mobility model are also presented in Figs.~\ref{fig-electrons1} and \ref{fig-holes1}, and in the upper graphs of Figs.~\ref{mod_pars_1} and \ref{mod_pars_11}. They show, that the release of charge carriers from the plasma cloud can be significantly accelerated, when the diffusion parameter, which is related in our simulation to the mobility by the Einstein relation ($D_{n,p}=\mu_{n,p}k_B T/q$), does not decrease with increasing field. Combined with the increased drift velocity of the charge carriers this results in almost all cases in simulated current pulses, which are shorter than the measured ones. This observation suggests that the Einstein relation between mobility and diffusion, which is valid only in the equilibrium case, has to be modified at high fields and/or at high charge carrier densities.

The possible effects of Fermi-Dirac statistics (see \cite{BBK}, \cite{AGH})
were estimated and the expected influence on the pulse length should be studied, especially because the Einstein relation ($D=\mu k_B T/q$) would be replaced by a density dependent one. Hence the interplay of drift and diffusion will be modified.

The van Roosbroeck equations used in the simulations describe the situation in the diode for low density charge clouds well, however in situations with high densities gradients may violate assumptions made in the derivation of the drift-diffusion approximation (compare \cite{selberherr84} for a short discussion).

Optical photons of 660~nm wavelength (1.87~eV) produce so called hot charge carriers, as the band gap of silicon is approximately 1.12~eV. The simulation assumes charge carriers in thermal equilibrium with the crystal lattice, which is justified, as the thermalization of the hot carriers is usually very fast compared to the pulse duration. 



\section{Summary}
A simulation program was developed to model the transport of charge carriers for high densities in silicon sensors with emphasis on the impact on sensor performance of detectors for experiments at the European XFEL.

The numerical stability and applicability for sensor design purposes has been demonstrated.

As a result of the comparison of measurements and simulations it is concluded that the observed plasma effects cannot be described by using the mobility and diffusion models in literature. It is shown that, except for the highest intensity, two different sets of mobility models can be used to simulate pulses which are either systematically longer or systematically shorter than the measurements and thus allows to estimate the minimum and the maximum of the pulse duration.

In spite of the discussed discrepancies the simulation program is a valuable tool for the design and optimization of sensors and readout electronics for the European XFEL.

Combining the effects of Fermi-Dirac statistics in the diffusion process, variations in the parameters of the initial distribution of charge carriers and an optimized mobility model it may be possible to provide simulations which reproduce the measurements reasonably well for all intensities with a single, unified set of parameters.

\section{Acknowledgments}
We would like to thank Dr. E. Fretwurst for providing the sample.

This work was partly funded by the European XFEL, the Helmholtz Alliance ''Physics at the Terascale'' and the Federal Ministry of Education and Research.

\section{Appendix A}

\begin{table}[tb!]
\centering
\begin{tabular}{c|c}
Parameter                                       & Value\\
\hline
$\epsilon_r$                                    & 11.67\\
$\mu_n^0$                                       & 1448~cm$^2$/Vs \\
$\mu_p^0$                                       & 495~cm$^2$/Vs \\
$\alpha_n$                                      & 2.33\\
$\alpha_p$                                      & 2.23\\
$I$                                                            &  $n_i$\\
$C_{n}^{ref}$                   & 3$\times$10$^{16}$~cm$^{-3}$\\
$C_{p}^{ref}$                   & 4$\times$10$^{16}$~cm$^{-3}$\\
$S_{n}$                                                 & 350\\
$S_{p}$                                                 & 81\\
$N$                                                            & $\sum_i|C_i|$ \\ 
$\mu_{Adl1}$                            & $\frac{1.04 \times 10^{21}}{cm Vs}(\frac{T}{300K})^{3/2}$\\
$\mu_{Adl2}$                            & $\frac{7.45 \times 10^{13}}{cm^2} (\frac{T}{300K})^{2}$\\
$v_{sat_n}$                             &  1.1 $\times$ 10$^7$ cm/s\\
$v_{sat_p}$                             &  0.95 $\times$ 10$^7$ cm/s\\
$\beta_n$                                       & 1 or 2\\
$\beta_p$                                       & 1\\
\end{tabular}
\caption{Parameters used in the mobility parameterization. Parameters are quotes from \cite{selberherr84}, $\beta_n$ has been set to unity.}
\label{pars}
\end{table}

\begin{table}[tb!]
\centering
\begin{tabular}{c|c}
Parameter  & Value\\
\hline
$\mu_{p}^0(T)$ & $474 \;\frac{cm^2}{Vs} \;{T_{rel}}^{-2.619}$ \\
$v_{sat_p}(T)$ & $0.94 \;\times10^7 \frac{cm}{s} \;{T_{rel}}^{-0.226}$ \\
$\beta_p$(T) & $1.181  \;{T_{rel}}^{0.644}$ \\
\hline
$\mu_{n}^0(T)$ & $1440  \;\frac{cm^2}{Vs} \;{T_{rel}}^{-2.26}$ \\
$v_{sat_n}(T)$ & $1.054 \;\times10^7 \frac{cm}{s} \;{T_{rel}}^{-0.602}$ \\
$\beta_n(T)$ & $0.992 \;{T_{rel}}^{0.572}$
\end{tabular}
\caption{Parameters from \cite{becker2} used for the parameterization labeled 'fitted mobility'. The abbreviation $T_{rel}=\frac{T}{300K}$ has been used to improve legibility.}
\label{mobpar}
\end{table}

\subsection{Simulation parameters and derived values}
All symbols and their meaning are listed in Table \ref{pars2}. All potentials in this table are normalized to a constant reference potential ($U_T=k_B T_0 /q \approx 1/40 V$ at room temperature), all densities to a reference density ($n_{ref}=10^{10} {\rm cm}^{-3}\approx n_i$).

\subsection{Assumptions on the simulation domain, grids, and space discretization}
The simulation space is defined on a bounded, polyhedral domain $\Omega \subset I\!\!R^3$ with a boundary $\partial \Omega  = \Gamma_D \cup \Gamma_N$. $\Gamma_D$ denotes the Dirichlet part of the boundary. It is closed, has a positive measure and describes ohmic contacts. The Neumann part $\Gamma_N$ describes insulating boundary parts or symmetries. The solutions $(w(t),n(t),p(t))$ are defined in $S \times \Omega $ with the time interval $S =(0,t_{end})$ and $(w(0)=w_0, n(0)=n_0, p(0)=p_0)$ the initial values. 
The domain is discretized by a boundary conforming, tetrahedral Delaunay mesh and the usual Scharfetter--Gummel discretization is used (compare \cite{kg09}, \cite{hs05}).

\begin{table*}[tb!]
\centering
\begin{tabular}{l|l}
Parameter                                       & Meaning\\
\hline
$t$													& time\\
$x$ 												& position in space (vector)\\
$\epsilon_0$ 								& dielectric constant\\ 
$\epsilon=\epsilon_0 \epsilon_r$ 									& dielectric permittivity\\ 
$k_B$												& Boltzmann constant\\
$q$													& elementary charge\\
$m_0$												&	electron mass\\
$m^*_{n,p}$									& effective mass of electron or hole\\
$a_{Bor}$										& Bohr radius\\
$\hbar$											& reduced Planck constant\\
$T$													& lattice temperature\\
$T_0=293.15K$								& reference temperature\\
$w$ 												& electrostatic potential\\ 
$\phi_n$ $=w - \log{n/n_i}$ 		& quasi-Fermi potential for electrons\\ 
$\phi_p$ $=w + \log{p/n_i}$ 		& quasi-Fermi potential for holes\\ 
$n=n_i e^{w-\phi_n}$						& electron density\\
$p=n_i e^{\phi_p-w}$  					& hole density\\
$C$ 												& density of impurities\\
$\tau_n=\tau_p=10^{-3}$s		& Shockley-Read-Hall recombination lifetime\\
$p_0=n_0=n_i$								& intrinsic charge carrier density\\
$R=\frac{1}{\tau_n p_0+\tau_p n_0+\tau_n p+\tau_p n}(n_i^2-n p)$& recombination / generation rate\\
$\mu_{n,p} > 0$							& carrier mobilities \\
$D_{n,p}=\mu_{n,p}k_B T/q$	& Einstein relation\\ 
\end{tabular}
\caption{Parameters and derived values used in the simulations.}
\label{pars2}
\end{table*}

\subsection{Time integration}
For the implicit Euler scheme dissipativity can be pro\-ven. It guarantees
positive solutions on any Delaunay grid and for any time step. This cannot be expected for higher order schemes in general. Hence
the related family of backward differentiation methods (BDF) is used to reduce dissipation while the order control handles the observed oscillations in space and time.
Let
\[
y^\prime(t)=f(t,y(t)), ~~ y(t_0)=y_0
\]
denote a nonlinear first order initial value problem with the initial value $y_0$.
The BDF formulas for order $m=1,2,3$ ($m=1$ implicit Euler) at time step
$y(t_k)=y_k$ with constant step size $\tau$ are:
\begin{align}
& m=1: \; y_{k+1}-y_{k}=\tau f(t_{k+1},y_{k+1}) \\
& m=2: \; \frac{3}{2}y_{k+1}-2 y_{k}+ \frac{1}{2}y_{k-1} =\tau f(t_{k+1},y_{k+1})\\
& m=3: \; \frac{11}{6} y_{k+1} - 3 y_k + \frac{3}{2} y_{k-1} - \frac{1}{3} y_{k-2} =\tau f(t_{k+1},y_{k+1})
\end{align}

For $m=1, 2$ the formulas are A-stable (stable for the linear test problem and eigenvalues in the left half plane, the cases $m=3, 4$ are still stable for a sector of 88$^\circ$, 72$^\circ$ around the negative real axes), see \cite{bronstein} and literature cited therein.
The BDF formulas can be modified for variable step size.
Time step size control is based on predictor-corrector differences of
functionals, like free energy, dissipation rate, one selected contact current, 
sources, and their deviations from a polynomial predictor and the number 
of Newton steps needed to solve the nonlinear equations.
The variable time integration order is controlled by testing different order
predictors against the computed solution at the present time step. The order
of integration of the next time step is defined by the predictor with minimal
error.
Time step
rejections are based on the local truncation errors and the number of Newton 
steps. A rejection is combined with order reduction. The maximum order can be 
specified.
The identical technique is used to apply continuation methods (the continuation
parameter ($\tilde t$) replaces time ($t$)) with respect to the boundary values or model
parameters, because it is often impossible to reach large applied voltages
directly from the uniquely defined equilibrium. $\tilde t=0$  corresponds to a
state with known solution, $\tilde t=1$ to the wanted state with an unknown solution.
The only difference, with respect to time integration, is taking truncation
errors not into account, because a sequence of stationary solutions is constructed.
The predictor is chosen such that the positivity of the density variables is
preserved.

Each time step requires the precise solution of one nonlinear system of
equations by Newton's method. Charge conservation is lost accordingly due to the
errors introduced in the solution of the nonlinear equations.

\subsection{Solution of the linear and nonlinear equations}
The discrete nonlinear system of equations is Newton linearized by computing the functions and their derivatives of all dependencies, including the derivatives of parameters, together.
Hence, common expressions are reused in the function and the Jacobian.
The solution of the nonlinear equations is controlled by the $L^\infty$-norm 
of all potential updates, hence it is precise for low densities, too. 

The linear systems are solved by a combination of direct \cite{oschenk:03}, iterative methods \cite{cgs} and primary and secondary preconditioners. The primary 
preconditioner avoids the  factorization of the complete Jacobian by approximating
the main dependencies by scalar equations on the whole domain. The 
secondary preconditioners solve local systems to include the missing couplings
(avalanche, small time steps, strong recombination etc.) for all variables at
a given grid point or very small subsets of the grid. 

This allows to avoid
the factorization of the complete Jacobian, hence to reduce memory requirements
and the operations needed while one is still getting quadratic convergence of the Newton process.

The main algorithms defining the limits of the applications are grid generation
and the solution of the linear systems.

\end{document}